\documentclass[3p,times,twocolumn]{elsarticle}

\usepackage{sqrcaps}
\usepackage[T1]{fontenc}

\usepackage{ulem}

\usepackage{graphicx}
\usepackage{dcolumn}
\usepackage{bm}
\usepackage{epstopdf}

 \usepackage{lineno}
 \usepackage{amssymb}

 \newcommand{\bD}{{\usefont{U}{dutchcal}{b}{n}D}}
  \newcommand{\bO}{{\usefont{U}{dutchcal}{b}{n}O}} 
 \newcommand{\bS}{\underline{\usefont{U}{eus}{b}{n}S\normalfont}}
\newcommand{\bDhat}{\underline{{\bD}$^{\hspace{-1mm}\hat{}}$}}
\newcommand{\bOhat}{\underline{{\bO}$^{\hspace{-1mm}\hat{}}$}}

\newcommand{\up}{\uparrow}
\newcommand{\down}{\downarrow}

\newcommand{\br}{{\it h}}
\newcommand{\myskip}[1]{}   
\newcommand{\mijnskip}[1]{}   

\newcommand{\BEQ}{\begin{eqnarray}}      
\newcommand{\EEQ}{\end{eqnarray}}      
\newcommand{\BEA}{\begin{eqnarray}}      
\newcommand{\EEA}{\end{eqnarray}}      
\newcommand{\nn}{\nonumber }      
\renewcommand{\d}{{\rm d}}      
\newcommand{\half}{\frac{1}{2}}

\newcommand{\dec}{{\rm dec}}
\newcommand{\off}{{\rm off}}

\newcommand{\SA}{{\rm SA}}

\newcommand{\scriptD}{\hat{\cal D}}
\newcommand{\scriptE}{{\cal E}}
\newcommand{\scriptH}{{\cal H}}
\newcommand{\scriptN}{{\cal N}}
\newcommand{\scriptR}{\hat{\cal R}}
\newcommand{\tf}{t_{\rm f}}
\newcommand{\tr}{{\rm tr}}

\newcommand{\sub}{{\rm sub}}
\newcommand{\eq}{{\rm eq}}

\newcommand{\rS}{{\rm S}}

\newcommand{\tpf}{{t'_{\rm f}}}
\newcommand{\tsplit}{\tpf}

\newcommand{\shr}{{\rm shr}}

\begin{document}

\title{A sub-ensemble theory of ideal quantum measurement processes }
 
 \author{Armen E. Allahverdyan$^1$, Roger Balian$^2$ and Theo M. Nieuwenhuizen$^{3,4}$}

\address{$^1$ Yerevan Physics Institute, Alikhanian Brothers Street 2, Yerevan 375036, Armenia \\
$^2$  Institut de Physique Th\'eorique, CEA Saclay, 91191 Gif-sur-Yvette  cedex, France \\
  $^3$ Institute for Theoretical Physics, 
Science Park 904, 
1090 GL Amsterdam, The Netherlands \\
$^4$  International Institute of Physics, Lagoa Nova 59078-970, CP: 1613 
 -  Natal/RN, Brazil }

\begin{abstract}
In order to elucidate the properties currently attributed to ideal measurements, one must explain how the concept of an individual event with a well-defined outcome may 
emerge from quantum theory which deals with statistical ensembles, and how different runs issued from the same initial state may end up with different final states. 
This so-called ``measurement problem'' is tackled with two guidelines. On the one hand,  the dynamics of the macroscopic apparatus A coupled to the tested system 
S is described mathematically within a standard quantum formalism, where ``q-probablities''  remain devoid of interpretation.
On the other hand, interpretative principles, aimed to be minimal, are introduced  to account for the expected features of ideal measurements. 
Most of the five principles stated here, which relate the quantum formalism to physical reality, are straightforward and refer to macroscopic variables. 
The process can be identified with a relaxation of S+A to thermodynamic equilibrium, not only for a large ensemble $\scriptE$ of runs but even for its sub-ensembles. 
The different mechanisms of quantum statistical dynamics that ensure these types of relaxation are exhibited, and the required properties of the Hamiltonian 
of S+A are indicated. 
The additional theoretical information provided by the study of sub-ensembles remove Schr\"odinger's quantum ambiguity of the final density 
operator for $\scriptE$ which hinders its direct interpretation, and bring out a commutative behaviour of the pointer observable at the final time. 
The latter property supports the introduction of a last interpretative principle, needed to switch from the statistical ensembles and sub-ensembles described 
by quantum theory to individual experimental events. It amounts to identify some formal ``q-probabilities'' with ordinary frequencies, but only those which refer to the final indications 
of the pointer. 
The desired properties of ideal measurements, in particular the uniqueness of the result for each individual run of the ensemble and von Neumann's reduction, 
are thereby recovered with economic interpretations. The status of Born's rule involving both A and S is re-evaluated, and contextuality of quantum measurements is made obvious.
\end{abstract}

 \maketitle




DOI:  {Subject Areas: Quantum Physics}
 
\section{Introduction}\label{sec1}


\hangindent=0.5cm

 {\it If one wants to be clear about what is meant by
 
   ``position of an object'', for example of an electron...,    
  
     then one has to specify definite  experiments by      
 
      which the ``position of an electron'' can be measured;  
 
       otherwise this term has no meaning at all.}

\hfill{Werner Heisenberg \cite{heisenberg1925uber}}

\hspace{3mm}


  Measurements constitute our sole contact with microscopic reality, but raise many questions, closely related to the connection between microscopic and 
 macroscopic concepts. Can one explain theoretically why identical measurements performed on several systems identically prepared provide different 
 outcomes? For a single measurement, how is the occurrence of a well-defined result compatible with the irreducibly probabilistic nature of quantum theory? 
 Does measurement theory require a specific principle of quantum mechanics? What is the status of Born's rule? What is the role of the apparatus? 
 Already raised by the founding fathers, these questions have witnessed a revival 
 \cite{wheeler2014quantum,alter2001quantum,braginsky1995quantum,home1992ensemble,namiki1993quantum,zurek2003decoherence,
 sewell2007can,laloe2012we,allahverdyan2013understanding,narnhofer2014reduction,weinberg2016happens,demuynck2006foundations,demuynck2016crucial}.
 Many answers have been proposed, relying on various interpretations or on various extensions of quantum mechanics and often inspired by the solution of models, 
 but no consensus has been reached.

Here, we regard as usual a measurement as a joint process undergone by the tested system S and a macroscopic apparatus A; the dynamics of a large 
statistical ensemble $\scriptE$ of similarly prepared runs is represented by current equations of quantum statistical mechanics. The mathematical results thereby 
obtained must then be interpreted, so as to relate them to physical facts. However, we do not wish to adopt any specific interpretation of the quantum formalism. 
Our purpose is more modest, as we will limit ourselves to the interpretation of the sole results relevant to the measurement, keeping aside all other quantum 
degrees of freedom. 

Our scope is thus double, {\it technical} and {\it conceptual}. On the technical side, we wish to sort out how much can be told about ideal measurements 
through a quantum approach restricted to a formal skeleton devoid of any interpretation. To this aim, we will study the dynamics of S+A, not only in the standard way 
where the density operator describes the whole ensemble $\scriptE$ of realisations of the experiment, but also by introducing more precise density operators which 
describe {\it sub-ensembles} of $\scriptE$. Governed by the same conventional equations of motion, these  operators will provide some useful, more detailed information. 

On the conceptual side, in order to explain the various features expected for ideal measurements, we will introduce {\it interpretative principles} that link some {\it formal} 
outcomes of the quantum analysis to the {\it physical} facts pertaining to the measurement. Discussions about interpretation will require a clear distinction between 
abstract quantum ``probabilities'' (termed {\it q-probabilities}) and ordinary probabilities regarded as frequencies of observing some macroscopic events in the limit 
of a large number of repeated experiments.  As we wish to introduce only 
the {\it most economic} principles (or postulates) needed to understand ideal measurements, most quantum objects will be left without interpretation. 
Several points below will appear well known or trivial; we included them for completeness and continuity of the reasoning.

\subsection {The measurement problem} \label{subsec1.1}

Let us pose more precisely the problem to be solved and define the notations. We deal with ideal, non demolishing  measurements. 
Their purpose is to test a single observable $\hat s=\sum_i s_i \hat\pi_i$ of S (characterised by its (discrete) eigenvalues $s_i$ and the 
associated eigenprojectors $\hat\pi_i$), while perturbing S minimally.  
 For instance, in the historical experiment of Stern and Gerlach (1922), the system S is one among the silver atoms of a beam incoming along the $x$-direction, 
it is coupled to the apparatus through an inhomogeneous magnetic field in the $z$-direction; the tested observable $\hat s$ is then the $z$-component of the spin of S, 
and the projectors $\hat\pi_i$ refer to the directions $+z$ and $-z$. 
 In EPR settings, $\hat\pi_i$ denotes the product of two projectors pertaining to the two correlated spins. 

As measurements are required to provide experimental access to microscopic physical quantities, their understanding is an essential element to settle the 
foundations and the interpretation of quantum mechanics.
Although the conditions of ideality recalled below are rarely fulfilled in the laboratory, it is natural to focus as we do here on the simplest case of ideal measurements. 
Indeed, only ideal measurements are dealt with in quantum mechanics textbooks, which postulate their characteristic properties 
(Born's rule and von Neumann's reduction) but skip the analysis of the quantum process of interaction between the tested system and the apparatus,
 needed to justify these postulates. Moreover, as any general quantum measurement (POVM) can be represented as a partial trace over an ideal measurement
\cite{demuynck2006foundations}, a theoretical elucidation of ideal measurements appears as a prerequisite for a full understanding of real measurements, 
which should rely on the same ideas.

 An essential feature is the {\it macroscopic size} of the apparatus A, which forces us to deal with mixed states and non-equilibrium quantum statistical mechanics. 
 We denote by $\scriptD(t)$ the joint density operator of S+A for a large ensemble $\scriptE$ of runs, 
 and  by  $\hat  r(t)={\rm tr}_{\rm A} \scriptD(t)$ and $\scriptR(t)={\rm tr}_{\rm S} \scriptD(t)$ the marginal density operators of S and A, respectively. 
 At the initial time $t=0$, S and A are uncorrelated, S lies in some state $\hat r(0)$, pure or not, to be tested and A lies in a {\it metastable state}\footnote{This 
 initial state of A is often called ``ready state'', waiting to be triggered by S. It must therefore be metastable, and hence can be represented only by a mixed 
 density operator $\scriptR(0)$, not by a pure state, a property often overlooked.}  $\scriptR(0)$, so that $\scriptD(0)$ equals $\hat r(0)\otimes\scriptR(0)$.

The subsequent evolution of S+A should obey quantum statistical dynamics. We expect that the apparatus, triggered by an interaction $\hat H_{\rm SA}$ 
 with S which is first switched on and later off, will eventually relax towards one or another among its {\it stable states} $\scriptR_i$. These states should have 
 equal entropies and energies so as to avoid bias in the measurement; they can be distinguished from one another by observing, processing or registering the 
 value $A_i$ of the pointer variable, identified as the expectation value $A_i =$ tr$_{\rm A} \scriptR_i \hat A$ in the state $\scriptR_i$ of some collective 
 observable $\hat A$ of A. As the pointer is macroscopic, the spectrum of $\hat A$ is dense, and many eigenvalues of $\hat A$ lie in the range of each 
 distribution tr$_{\rm A}\scriptR_i \delta(A-\hat A)$. Moreover, these distributions should not overlap for $i\neq j$ so as to ensure a neat distinction between 
 the possible outcomes $A_i$. Introducing a width $\Delta$ larger than that of $\scriptR_i$ and such that $\Delta\ll |A_i - A_j|$ for $i\neq j$, we shall denote 
 as $\hat\Pi_i$ the projector\footnote{It is essential to distinguish the projector $\hat \pi_i$ for the system S, associated with the eigenvalue $s_i$ of $\hat s$, 
 from the projector $\hat\Pi_i$ for the apparatus A, associated with the eigenvalues of $\hat A$ located in the range 
 ($A_i-\Delta, A_i+\Delta$).\label{newfootone}} on the eigenspace characterised by eigenvalues of $\hat A$ lying between 
 $A_i - \Delta$ and $A_i + \Delta$. We then have $\tr_{\rm A} \scriptR_i \hat\Pi_j \simeq  \delta_{ij} $. 

An ideal measurement of the tested observable $\hat s $, performed on the initial state $\hat r(0)$ of S, is currently defined as a thought experiment which is supposed 
to have the following properties. The experiment involves a large number of runs during which S and A interact. One assumes that these runs can be sorted out at the 
final time $\tf$ according to the macroscopic indication $A_i$ of the pointer, and that the relative number$^{\ref{newfootone}}$ tr $\scriptD(\tf)\hat\Pi_i$ of runs 
having yielded the outcome $A_i$ is given by Born's rule $p_i=\tr\,\scriptD(\tf)\hat\pi_i = \tr_{\rm S}\, \hat r(0)\hat\pi_i $. One also admits that the outcome $A_i$ is
 fully correlated with the eigenvalue $s_i$ of $\hat s$ and with the production of the final state $\hat r_i =\hat\pi_i \hat r(0)\hat\pi_i / p_i $ of S (von Neumann's reduction or
so-called collapse postulate, see L\"uders  \cite{luders1950zustandsanderung}\footnote{English translation and discussion: K. A. Kirkpatrick \cite{luders2006concerning}.}).
These properties are expressed by the surmise that {\it one among the final states} of the form 
 
\BEQ \label{fin-i}        
\scriptD_i = \hat r_i \otimes \scriptR_i, \qquad 
\label{ri}
\hat r_i =\frac{1}{p_i} \hat\pi_i \hat r(0) \hat\pi_i,
\EEQ
with $p_i= \tr_{\rm S}\, \hat r(0)\hat\pi_i $, should be assigned to S+A {\it after each separate run} of the measurement.

A major difficulty arises when one  tries to show that the above features traditionally attributed to ideal measurements result from the application of 
quantum theory to the dynamics of the compound system S+A. Indeed, the very definition of a measurement relies on the concept of {\it single run}, 
whereas this concept is foreign to standard quantum mechanics which only deals with {\it large statistical ensembles}. 
One is thus faced with the so-called {\it ``measurement problem''}  \cite{laloe2012we}. (For pure states a clear definition is given by Home \cite{home2013conceptual}.)
To solve it,  one must supplement the abstract formalism of quantum mechanics with some interpretative principles so as to give way to the concept of individual runs in 
  spite of the inevitably probabilistic nature of quantum theory. 
  The main principle that we propose (Principle 5, Sec. 6) will concern only the macroscopic (quantum) apparatus, not the microscopic tested system. 
  Afterwards, one may be in position $(i)$ to understand why each individual run produces a well-defined outcome $A_i$, ($ii$) to elucidate how 
  different runs issued from the {\it same initial state} $\scriptD(0)=\hat r(0) \otimes  \hat {\cal R}(0)$ through deterministic quantum equations of motion may 
  end up in {\it different final states} having the reduced form $\scriptD_i$, and ($iii$) to demonstrate why the {\it frequencies} of the pointer values $A_i$ 
  converge for a large ensemble of runs to Born's {\it formal q-probabilities} $p_i= \tr_{\rm S}\, \hat r(0)\hat\pi_i $ which refer only to the system 
  and to its initial state, irrespective of the apparatus and of the evolution.

\subsection {Outline: a sub-ensemble based approach} \label{subsec1.2} 

The explanation we wish to give to the desired properties of ideal measurements has been subjected to a double constraint. We tried to describe the dynamics 
of S+A by extracting as much mathematical results as possible from a standard quantum formalism, and at the same time to interpret the formal outcomes thus 
obtained in terms of physical reality by introducing the weakest possible postulates (or principles) needed. The two different types of ingredients that enter the 
approach, formal and conceptual, will be intertwined. In order to distinguish them, we exhibit all along the text five {\it ``interpretative principles''} which relate 
some mathematical objects to physical properties. Most of these principles concern {\it only macroscopic variables} through which we have access to reality. 

The mathematical formalism on which we rely, recalled in Sec. 2, is completed by the first three principles, which we state for completeness but may in fact be
 regarded as natural or evident. The {\it principle 1} (Subsec. 2.2) identifies the formal q-expectation value $\tr \,\scriptD\hat O$ of a macroscopic observable 
 $\hat O$ in state $\scriptD$ with the corresponding physical quantity, in case the corresponding q-variance of $\hat O$ is negligible. 
 The {\it principle 2} (Subsec. 2.3), relevant  for the dynamics of S+A, allows us to rely on approximations of quantum statistical mechanics that have negligible 
 effects upon the physical outcomes owing to the large size of A. The {\it principle 3} (Subsec. 3.2), which determines the density operator that should be assigned 
 to a system in a situation characterised by some data, is used here to interpret the expressions of the initial and final states of S+A.    

The quantum equations of motion refer to a large ensemble $\scriptE$ of compound systems S+A, but also apply to sub-ensembles of $\scriptE$ (Subsec. 2.4). 
 We therefore proceed in three steps, which involve successively ($i$) the full set of runs of the measurement, ($ii$) its sub-ensembles and ($iii$) the individual runs.   

\vspace{3mm}

{\it Step (i):  Full ensemble}. This step has commonly been worked out in the literature. The density operator $\scriptD(t)$ of the compound system S+A 
encodes the properties at the time $t$ of a large statistical ensemble $\scriptE$ of realisations of the measurement. 
 All elements of $\scriptE$ are initially prepared in an identical manner; the result is encoded by the state 
 $\scriptD(0)=\hat r(0)\otimes\scriptR(0)$ and the dynamics of $\scriptD(t)$ is governed by the Liouville--von Neumann equation. 
 One first needs to prove that, at the final time $\tf$ of the process, $\scriptD(t)$ reaches

 \BEQ
 \hspace{-0.7cm}
\scriptD(\tf) = \! \sum_i p_i \scriptD_i, \qquad  \scriptD_i=\hat r_i \otimes \scriptR_i,
\label{fin} 
\EEQ
which is a requirement needed for the desired result (1).
 
We will first identify $\scriptD(\tf)$ as a generalised Gibbs state (Subsec. \ref{subsec3.3}), so that the dynamics which leads from $\scriptD(0)$ to $\scriptD(\tf)$ 
 can merely be regarded as a {\it relaxation towards a thermodynamic equilibrium state}. We then show how this relaxation can be ensured dynamically within 
 a purely formal approach (Sec. \ref{sec4}), by deriving (2) through current methods of quantum statistical mechanics and by relying on some suitable 
 properties of the Hamiltonian of S+A (Subsecs. \ref{subsec3.1} and \ref{subsec3.4}). 

   What we wish to eventually demonstrate is that, after the final time $\tf$, the ensemble $\scriptE$ can be split into sub-ensembles $\scriptE_i$ characterised 
   by the macroscopic outcome $A_i$ of the pointer; for each sub-ensemble $\scriptE_i$, the compound system S+A should lie in the state $\scriptD_i$ given 
   by (1), and $\scriptE_i$ should contain a proportion $p_i$ of runs. The result (2) is a {\it necessary condition} for these properties to be satisfied, but 
   {\it one cannot ensure the converse} for quantum reasons. If density operators did behave as distributions of classical statistical mechanics, 
   one would be allowed to readily interpret each operator $\scriptD_i$ that enters (2) as a physical state and its coefficient $p_i$ as an ordinary probability. However, 
   Schr\"odinger's {\it quantum ambiguity} \cite{schrodinger1935discussion,schrodinger1936probability,park1988thermodynamic}, implies that the operator $\scriptD(\tf)$ can be decomposed not only into a weighted sum of operators $\scriptD_i$ as in (2), but also into very many other sums involving different terms.
 We shall recall (Subsec. 5.1) how contradictions arise when one attempts to interpret the separate terms of two different decompositions. 
 As nothing privileges a priori the decomposition suggested by the form of (2), the sole 
   establishment of this expression is {\it not sufficient} to ensure that {\it each of its separate terms} is physically meaningful. Other theoretical ingredients 
   will help us to find a natural interpretation for the components $\scriptD_i$ and $p_i$ of $\scriptD(\tf)$.
   
\vspace{3mm}

{\it Step  (ii): Sub-ensembles}. In order to draw further conclusions within the abstract formulation of quantum theory, we will take advantage of the fact 
   that quantum dynamics governs not only ensembles, but also sub-ensembles. We will therefore make an intermediate step, when going from the full 
   ensemble $\scriptE$ towards individual runs. We consider an arbitrary sub-ensemble $\scriptE_\sub^{(k)}$ 
   of runs extracted from $\scriptE$, which includes a proportion $q_i^{(k)}$ of runs having yielded the outcome $A_i$. We need to prove that the state of 
 S+A which describes $\scriptE_\sub^{(k)}$ ends up in the form

\BEQ	
    \scriptD_\sub^{(k)} (\tf)=\sum_i q_i^{(k)} \scriptD_i,\qquad  
\scriptD_i=\hat r_i \otimes \scriptR_i,
\label{fin-sub} 
\EEQ
with $0\le q_i^{(k)}\le 1$ and $\sum_i q_i^{(k)}=1$.
Contrary to what would happen in classical statistical physics, this expression (3) is not a consequence of (2), as discussed in Subsec. \ref{subsec5.1}. 
 It is a further {\it necessary condition, much stronger than} (2), and it must really be demonstrated.
 
Here again as for (2), the desired density operator (3) expresses thermodynamic equilibrium (Subsec. \ref{subsec3.3}). In order to give, within the standard 
 quantum formalism, a dynamical proof of the relaxation of S+A towards this expression (3) for the sub-ensemble $\scriptE^{(k)}_\sub$, 
  we introduce in Subsec. 5.2 the {\it principle 4}, which allows {\it under some conditions} to describe S+A in a {\it more precise} way than with $\scriptD(t)$ 
  by associating with the various sub-ensembles of $\scriptE$ different quantum states. 
 We thereby assume that, at least after some time $\tpf$ slightly earlier than $\tf$, 
 the dynamics of a physical sub-ensemble $\scriptE^{(k)}_\sub$ of $\scriptE$ is generated by ordinary quantum equations, even though our available information 
 is not sufficient to fully specify the state $\scriptD^{(k)}_\sub (\tpf)$ of S+A describing $\scriptE^{(k)}_\sub$ at the time $\tpf$. Then, making use of a specific 
 dynamical mechanism, the {\it ``poly-microcanonical relaxation''} (introduced in  \cite{allahverdyan2013understanding} under the name of ``sub-ensemble relaxation'') 
 which involves {\it only the (large) apparatus}, we can establish for any physical sub-ensemble the expected result (3), 
 thus {\it removing the quantum ambiguity} (Subsec. \ref{subsec5.4}).

\vspace{3mm}

{\it Step (iii): Individual runs.} The result (\ref{fin-sub}), much stronger than (\ref{fin}), is the {\it most detailed property} of ideal measurements that conventional 
 quantum theory can afford. It is a necessary condition, but its mere derivation is not sufficient to entail (\ref{fin-i}), because individual runs lie beyond the realm 
 of the standard formulation of quantum mechanics, and because the ingredients $\scriptD_i$, $p_i$ and $q_i^{(k)}$ of (2) and (3) are still formal 
 quantum quantities. Since no interpretation has yet been given to q-probabilities, the numbers $q_i^{(k)}$ entering (3) are only mathematical objects, 
 which we indeed would like to interpret as ordinary probabilities.
  
We will therefore supplement (Subsec. 6.1)  the abstract formulation of quantum mechanics with a last {\it principle 5}.  Its introduction is 
 made natural by the classical-like properties of the projectors $\hat\Pi_i$ in the final state (Subsec. 5.3), which result from the {\it macroscopic size of the pointer}
 and from the dynamics. It amounts to interpret, for any sub-ensemble $\scriptE_\sub^{(k)}$, each formal coefficient $q_i^{(k)}$ as the 
 {\it proportion of runs of $\scriptE_\sub^{(k)}$ that provide the indication $A_i$ of the pointer}. 
Equivalently, it amounts to acknowledge the existence of the sub-ensembles $\scriptE_i$ characterised by the value $A_i$ 
(for which $q^{(k)}_i=1$ and $q^{(k)}_{i'}=0$ for $i'\neq i$). 
 Accordingly the building block $\scriptD_i$  of the formal expressions (2) and (3) is identified with the final state associated 
 with the sub-ensemble $\scriptE_i$, so that it can be assigned to S+A for any individual run of $\scriptE_i$. Statements can thus be made
 about experimental facts, and all expected properties of ideal measurements come out (Sec. \ref{sec6}). 

We will stress in the conclusion (Sec. \ref{sec7}) that the features of ideal measurements emerge owing to the macroscopic size of the {\it apparatus}, 
which plays a major role in the interpretation. Accordingly, results of measurements involving different settings of apparatuses 
should not be put together (Subsec. 7.4). We will also reconsider Born's rule as a property of the apparatus in the final state after its interaction with S (Subsec. 6.4).

\vspace{3mm}
 
 The formal aspects of the theory lie in the derivation of Eqs. (2) and (3). Such derivations have been achieved at least partly for many specific
 models  \cite{ sewell2007can,laloe2012we,allahverdyan2013understanding,narnhofer2014reduction}.  As we consider below general ideal measurements,
 we will simply sketch how the solution arises from some necessary properties of the Hamiltonian of S+A, and demonstrate its technical feasibility by recalling 
 in footnotes the main features of the detailed dynamical study \cite{allahverdyan2013understanding,allahverdyan2003curie} of the Curie--Weiss 
 (CW) model of quantum measurement\footnote{In the CW model (see ref. \cite{allahverdyan2013understanding}, sect. 3), 
 S is a spin $\half$, the measured observable being its $z$-component  $\hat s_z$, with outcomes $ i=\,\up$ or $\down$. 
 The apparatus simulates a magnetic dot, including $\scriptN\gg 1$ spins
$\mathbf{\hat{\sigma}}^{\left(  n\right)}$, which interact through the Ising coupling 
$J$, and a phonon thermal bath at temperature $T<J$; these spins and the phonons are coupled through a dimensionless weak coupling $\gamma$. 
 Initially prepared in its metastable paramagnetic state,  A may switch to one or the other stable ferro\-magnetic state. 
The pointer observable $\hat A = N \hat m=\sum_{n=1} ^N  \hat \sigma_z ^{(n)}$ is the total magnetisation in the $z$-direction of the $N$ Ising spins. The coupling 
 between S and A is $\hat H_{\rm SA}= - \sum_{n=1} ^N  g \hat s_z \hat\sigma_z ^{(n)}$, while $\hat H_{\rm S}=0$. \label{footCW}}. 
 
 Moreover, since the derivation 
 of Eqs. (2) and (3) merely amounts to a proof, in the microscopic framework of quantum statistical dynamics, of the relaxation of S+A towards thermodynamic 
 equilibrium (Subsec. \ref{subsec3.4}), the reader willing to admit this thermalization may skip Secs. \ref{sec4} and \ref{sec5}.

\section{Formal principles of quantum mechanics} \label{sec2}

We tackle the measurement problem within a formulation of quantum mechanics which deals only with statistical ensembles. 
Indeed, this idea underlies most current interpretations of quantum mechanics, and repeated experiments constitute an exploration of the considered ensemble. 
 Individual systems are not directly described in this framework, which is irreducibly probabilistic, so that statistical ensembles and sub-ensembles will 
 be essential in our approach\footnote{We do not allude here to ``statistical interpretation'' \cite{home1992ensemble} nor to ``ensemble interpretation'', 
 terms which depend on the authors, but simply to ``formulation'' because interpretation will come out only in the end as a result of a measurement process. 
 We shall abbreviate throughout by ``ensemble'' the expression ``statistical ensemble''.}. The spirit of this formal description is the same as in the $C^\ast$-algebraic 
 approach  \cite{narnhofer2014reduction,thirring2013course,david2015quantum}, although we deal 
 here with finite non relativistic systems. Its principles recalled below do not prejudge any specific interpretation of quantum oddities \cite{demuynck2006foundations}, and it is 
 suited to both microscopic and macroscopic systems. In fact, S is microscopic and the macroscopic apparatus A is treated as a {\it finite} (though large) object 
 so as to keep control of the time scales characterizing the evolution of S+A. 
  
\subsection{The abstract formalism} \label{subsec2.1}

Physical quantities pertaining to a system are represented by {\it ``observables''} expressed as Hermitean matrices in a Hilbert space. Observables behave 
 as random objects, but, unlike ordinary random variables, their randomness, which arises from their non commutative nature, is inherent to the quantum formalism.

In the present formal scope, we regard a {\it ``quantum state''}, whether pure or not, merely as a theoretical tool for making probabilistic statements or predictions 
 about experiments\footnote{We subscribe to van Kampen's theorem IV on quantum measurements \cite{vankampen1988ten}, generalised  from pure states $\psi$ to general 
 mixed states $\scriptD$: ``Whoever endows $\scriptD$ with more meaning than is needed for computing observable phenomena is responsible for the consequences''.\label{vankampen1988tenIV}}. It is characterised by a correspondence that associates with any observable $\hat O$ a real number $\langle\hat O\rangle$\footnote{In this 
 algebraic approach, the observables $\hat O$ are regarded as elements of a vector space, while a state, defined as a linear correspondence 
 $\hat O\mapsto\langle\hat O\rangle$, such that $\langle\hat O\rangle$ is real and $\langle\hat O^2\rangle$ is non-negative, is an element of its dual vector space; 
 the q-expectation values $\langle\hat O\rangle$ appear as scalar products. The representation of states by density matrices arises when one chooses a set 
 of dyadics $|\eta\rangle\langle\eta'|$ as basis in the vector space of observables $\hat O$, which then appear as linear combinations of operators  
 $|\eta\rangle\langle\eta'|$ with coefficients $\langle\eta|\hat O|\eta'\rangle$. The matrix element $\langle\eta'|\scriptD|\eta\rangle$ of $\scriptD$ is then defined 
 as the q-expectation value of  $|\eta\rangle\langle\eta'|$.  Other so-called Liouville representations of states, such as the Wigner representation for a particle 
 or the polarisation representation for a spin $\frac{1}{2}$, are defined through other choices of bases in the dual vector spaces of observables and states 
 (the basis of Pauli operators for the polarisation representation of a spin) \cite{balian1999incomplete}.}. This correspondence is implemented as 
 $\hat O\mapsto\langle\hat O\rangle=\tr \,\scriptD\hat O$ by means of a Hermitean, normalised and non-negative density operator $\scriptD$. 

Such definitions of observables and states look analogous to the corresponding ones in classical statistical mechanics, where physical quantities are represented 
 by functions of the (random) position and momentum variables, where a state is encoded by a density in phase space, and where expectation values are 
 expressed as integrals over their product. However, this similitude is only formal, since the numbers $\langle\hat O\rangle$ violate some properties of ordinary 
 expectation values, for instance Bell's inequalities. Our knowledge is limited by the operator nature of the quantum physical quantities (and not only by some 
 ignorance about their values as in classical statistical mechanics). In particular, q-bits represented by two-by-two density matrices differ from ordinary bits. 
 They can be manipulated only blindly, since the ``quantum information'' (q-information) that they carry is not fully available: Reading a q-bit so as to extract 
 from it ordinary information in the form of a bit requires a measurement process which destroys it in part. Similarly, for a general density operator, the numbers 
 $\langle\hat O\rangle$ may become physically available in the form of ordinary expectation values solely in special circumstances and solely in part, 
 through measurements. 

One should therefore, as done for q-bits, distinguish $\langle\hat O\rangle=\tr \,\scriptD\hat O$ from an ordinary expectation value by denominating it as a 
 {\it ``q-expectation value''}. Likewise, a {\it ``q-correlation''}, the q-expectation value of a product of two observables, should  not be confused with an 
 ordinary correlation. Also, the q-expectation value $\langle\hat\pi\rangle$ of a projection operator 
 $\hat\pi$ is not an ordinary probability, but a formal object which we will call {\it ``q-probability''} rather than ``probability''\footnote{The term pre-probability has also 
 been proposed to indicate that the formal quantum object $p_i = \langle\hat\pi_i\rangle={\rm tr}_{\rm S} \hat r(0) \hat\pi_i$ may be interpreted as a true probability 
 only after achievement of an ideal measurement of $\hat s$.}. Born's rule is not postulated here, it will come out (Subsec. 6.4) as a property of the apparatus 
 at the issue of an ideal measurement.

 \subsection{On the threshold of interpretation}\label{subsec2.2}

 We want to extract from the abstract q-information embedded in density operators some 
 ordinary information affording predictions about real events. To this aim, physical interpretations should emerge at the macroscopic scale, in experimental contexts. 
 Let us already point out, for a macroscopic quantum system, a simple situation in which an interpretation is readily provided by a first, trivial principle.

\vspace{3mm}

{\it Interpretative principle 1}. If the q-variance of a {\it macroscopic observable} is negligible in relative size\footnote{This principle does not 
at all mean ``microscopic definiteness'' where the system is close to an eigenstate \cite{squires1990alleged,aharonov2002macroscopic,finkelstein2003comment}; we refrain from interpreting
microscopic properties.
Its use may in particular require the assignment of a lower bound to the q-variance of the considered macroscopic observable. 
See  footnote \ref{fnAppA} in Appendix A and ref.  \cite{balian1987equiprobability}.}, 
its q-expectation value is identified with the value  of the corresponding macroscopic physical variable, even for an individual system.    

\vspace{3mm}

Accordingly, the q-expectation value of $\hat A$  in the quantum state $\scriptR_i$ is identified with the macroscopic pointer value $A_i$. Nevertheless, in the 
 state $\scriptD(\tf)$ (Eq. (2)), the q-variance of $\hat A$ is in general large because its possible values $A_i$ are different,
 and the interpretation of $ \tr\scriptD(\tf)\hat A$ as an ordinary expectation value will 
 only arise from the analysis of the ideal measurement process of $\hat s$ and from the additional interpretative principle 5 (Sec. 6).

In spite of the macroscopic nature of the above principle, it can be used to provide a (somewhat roundabout) interpretation of q-expectation values, even for 
 microscopic systems (Appendix A and Ref. \cite{balian1987equiprobability}). Let us associate with the system S under study a macroscopic thought {\it super-system} 
 \bS {}\, =  $\{\rS^{[1]} , \rS^{[2]},\cdots , \rS^{[\scriptN]}\}$. It is a single compound system obtained by putting together a large number $\scriptN$ of subsystems 
 $\rS^{[n]}$ ($n=1, 2,\cdots, \scriptN$) similar to S. All these subsystems lie in the same marginal state $\scriptD$, obtained by tracing out the $\scriptN $ - 1 
 other subsystems from the state \bDhat \,\, of \bS.With each observable $\hat O$ of S we associate the average observable 
 \bOhat \, = $\scriptN^{-1}\sum_n \hat O^{[n]}$ of \bS. It is shown in Appendix A that, while the q-expectation values tr $\scriptD\hat {O}$ for S and 
 \uline{Tr} \bDhat\,\,\bOhat {}\,  \, for \bS \, are the same, the q-variance of \bOhat \,\, is $\scriptN$ times smaller than the q-variance of $\hat O$ 
 (provided the subsystems $\rS^{[n]}$ are sufficiently weakly q-correlated). The above principle thus holds for the macroscopic observable \bOhat \,, so that 
 the {\it formal q-expectation value} $\langle\hat O\rangle$ {\it for the (possibly small) system} S can be identified with the {\it macroscopic value of the 
 corresponding average observable} \bOhat \,\, for the super-system \bS. However, q-expectation values will remain without direct interpretation in terms of S itself.

\subsection{Dynamics} \label{subsec2.3}
 
The formalism is completed, for the time-dependence of the density operator of an isolated system, by the Liouville--von Neumann equation of motion 
 $i\hbar \d \scriptD(t)/\d t=[\hat H, \scriptD(t)]$. Mathematically, this fundamental dynamic equation is deterministic and reversible, whereas a measurement 
 process leading from $\scriptD(0)$ to $\scriptD(\tf)$ is irreversible. We thus have to face in this context the old {\it paradox of irreversibility}, like in classical 
 statistical mechanics, within replacement of the Liouville theorem in phase space by the unitarity in Hilbert space, and to solve it in the same way. 

As usual in statistical mechanics, it is legitimate in practice for finite but large systems to disregard events that might occur with an extremely {\it small probability}, 
 to forget about {\it recurrences} that might take place after large, unattainable times, and to neglect {\it physically irrelevant correlations} between a macroscopic 
 number of degrees of freedom. This view is consistent with the idea that a quantum state is regarded only as a catalogue of knowledge intended for physical predictions.
  Its evolution appears as a transfer of q-information among the various observables, the most complicated of which cannot be reached experimentally. 
 A part of the catalogue thus becomes useless and may be discarded (dissipation). Such a coarse graining breaks the constancy of entropy, replacing the 
 conserved von Neumann entropy by an increasing {\it relevant entropy} \cite{ balian2006microphysics}. We are led to the following prescription.

\vspace{3mm}

{\it Interpretative principle 2}. One may perform a coarse graining on a density operator ${\scriptD}$ if this operation has no effect on the physical predictions 
 afforded by ${\scriptD}$.

\vspace{3mm}

Standard  procedures in quantum statistical mechanics are thereby justified. For instance, correlations with a bath or an environment which develop during 
 the relaxation process are inaccessible and ineffective; they may be discarded. Such approximations, although not mathematically rigorous, are fully justified 
 when their outcome is physically indistinguishable from the exact solution. Moreover, they are necessary to explain irreversible phenomena, 
 including measurement processes.
 
 \subsection{Ensembles and sub-ensembles} \label{subsec2.4}
 
As an ordinary probability distribution, a quantum state gathering q-information refers, implicitly or not, to a statistical ensemble $\scriptE$, which is a large 
 collection of systems produced under the same conditions and characterised by the same available knowledge. However, while ordinary probabilities are 
 defined in terms of the individual events embedded in $\scriptE$, q-probabilities are abstract numbers which do not arise from the consideration of individual systems. 
 A ``state'' does not ``belong to a system'', it is not an intrinsic property but rather a catalogue of knowledge about an ensemble \cite{schrodinger1935discussion,schrodinger1936probability}.
  If one wishes to consider a single system, one should introduce a {\it virtual ensemble} $\scriptE$
 encompassing many mental copies of the studied system. However, a measurement gathers a large set of runs, and involves a {\it real ensemble} $\scriptE$ 
 of systems S+A, similarly prepared and evolving in repeated experiments. 

Note that {\it different density operators} may simultaneously be ascribed to the same system, {\it depending on the ensemble in which it is embedded}, that is, 
 on the information available about it. This is standard in probability theory: When a dice is repeatedly thrown, the probability of the outcome ``3'' is $\frac{1}{6}$ 
 for the full set of runs; it is  $\frac{1}{3}$ when only the odd outcomes (``1'', ``3'', ``5'') are selected and the even ones discarded; it is  $\frac{1}{2}$ for a selection 
 of the middle ones (``3'' or ``4''), and 1 for the sub-ensemble containing only the outcome ``3''. Gaining knowledge about an individual system which is originally part 
 of $\scriptE$ leads to regard it as member of a sub-ensemble of $\scriptE$, and to modify its probabilistic description by assigning to it a new, more informative state. 
 Such an occurrence of different probability distributions for the same system, {\it depending on the q-information} retained about it, which is trivial in the dice example, 
 may look odd for quantum states, but it takes place as soon as some non-random selection is made among measurement outcomes. 
  Hence it should enter theoretical treatments;  indeed, it will be crucial in Sec. \ref{sec5}. 
  Once the existence of different sub-ensembles is granted, the corresponding states evolve in parallel. 

However, a specifically quantum difficulty arises (Subsec. \ref{subsec5.1}). Knowing solely a mixed state such as $\scriptD(\tf)$ does not allow to recognise 
 theoretically within it states that might describe the sub-ensembles of $\scriptE$, nor a fortiori states that might describe its individual samples 
 (although these are evidently distinguished experimentally in repeated processes)\footnote{In classical probability theory, the selection of the elements of a sub-ensemble  
 $\scriptE_\sub^{(k)}$  of $\scriptE$ is mathematically implemented \cite{vonmises1957probability,birnbaum1940properties} by numbering the events of $\scriptE$ with an index 
 $n$ and introducing a function $f^{(k)}(n)$ that may take two values, 0 if the element $n$ is discarded, 1 if it is selected. The only general condition imposed 
 on the function $f$ is that the sub-ensemble should become infinite, whenever the ensemble does. The actual construction of $f$ can (but need not) be related 
 to distinguishing theoretically the individual events, which is of course experimentally performed in quantum measurements, but which is allowed in quantum theory 
 only in special cases, such as at the issue of a measurement. This will be discussed in Secs. \ref{sec5} and \ref{sec6}. Then, the sub-ensembles 
 $\scriptE_\sub^{(k)}$ of theoretical interest for the present argument will be those for which the coefficients $q_i^{(k)}$ in (3) differ from $p_i$. 
 However, the selections of all such sub-ensembles within the full ensemble  $\scriptE$ have zero measure (in the sense of Lebesgue measure in the space 
 of selections \cite{birnbaum1940properties} when the number of elements of $\scriptE$  becomes infinite). Nearly all subsets of $\scriptE$, in particular those 
 obtained by extracting systems at random from $\scriptE$, would be described by the same state $\scriptD$ as the full set $\scriptE$. It will therefore be 
 essential for our purpose to consider {\it all sub-ensembles} $\scriptE_\sub^{(k)}$ of $\scriptE$. Note also that, if $\hat r_i$ is a mixed state, the runs described 
 by (3) are picked up at random within $\scriptE_i$. 
 Note finally that, if we step away from measurements (for which the state of A is necessarily mixed) and consider a pure state $\scriptD=|\psi\rangle\langle\psi|$, 
 this same state $\scriptD$ should be assigned to any sub-ensemble and to any individual system of $\scriptE$. \label{newfn2}}. 
 The occurrence of $\scriptD_i$ within the expression (2) of $\scriptD(\tf)$ that describes the full ensemble $\scriptE$ is not sufficient to ensure that this operator 
 $\scriptD_i$ can be interpreted as final state assigned to an individual run, and we shall need both technical and conceptual developments to reach this conclusion. 

Indeed, the consideration of sub-ensembles, inspired from the frequency approach to the classical probability theory \cite{vonmises1957probability}, will be an essential ingredient 
 of the present approach to quantum measurements. While the density operator (2) of the compound system S+A encompasses q-information about the final state 
 of a large set $\scriptE$ of runs, the final states (\ref{fin-sub}) generated by some specific dynamics 
  (Subsec. \ref{subsec5.4}) will account for the more detailed q-information associated with the sub-ensembles $\scriptE_\sub^{(k)}$. 
 As for the final state $\scriptD_i$ of the form (1), understanding its occurrence requires solving the measurement problem, 
 as it is assigned to the individual runs of the sub-ensemble $\scriptE_i$ (Subsec. 6.1).  Switching from $\scriptD(\tf)$ to $\scriptD_i$ will appear as 
 an {\it updating} of information, similar to an updating associated with a gain of information, analogous to an updating of ordinary probabilities 
 after selection of events characterised by some piece of information.
 
\section{Preliminaries}\label{sec3}

\subsection{Hamiltonian generating an ideal measurement process}\label{subsec3.1}

Various measurement  models have been worked out 
\cite{heisenberg1925uber,wheeler2014quantum,alter2001quantum,braginsky1995quantum,home1992ensemble,namiki1993quantum,zurek2003decoherence, sewell2007can,laloe2012we,
allahverdyan2013understanding,narnhofer2014reduction,demuynck2006foundations,allahverdyan2003curie,vankampen1988ten,daneri1962quantum}. 
 In all cases, if we include in A a thermal bath or a possible environment, the compound system S+A is isolated, and therefore  governed by a Hamiltonian 
 $\hat H=\hat H_{\rm S}+\hat H_{\rm A}+\hat H_{\rm SA}$, which should ensure that the state of S+A evolves unitarily from $\scriptD(0)$ to $\scriptD(\tf)$. 
 (The same conclusion holds if the environment is left outside A, in which case the Liouville -- von Neumann evolution should be replaced by an equation 
 justified by quantum statistical mechanics.) 
 The most general Hamiltonian that may describe an ideal quantum measurement process should satisfy the following properties. 

The part $\hat H_{\rm A}$ associated with the macroscopic apparatus A alone must have specific features. It should produce an initial {\it metastable} state 
 $\scriptR(0)$, with lifetime longer than the duration of the measurement, and {\it several equilibrium} states $\scriptR_i$, the possible expected final states. 
 A typical example$^{\ref{footCW}}$ is given  by spontaneously broken discrete invariance, the macroscopic pointer variable $A_i$  being the order parameter 
 which may take two or more discrete values. These properties imply in particular the presence of a bath or an environment, coupled to the active part of A 
 including the pointer, which will drive it to thermodynamic equilibrium. The weakness of such a coupling allows to solve models by means of standard equations 
 of quantum statistical mechanics which eliminate the environment from the Hamiltonian dynamics.
 
As we wish to deal with ideal measurements, the process should perturb S as little as possible: any observable of S compatible with $\hat s$, i.e., commuting 
 with its eigenprojectors $\hat \pi_i$, should remain unaffected. The conservation of all these observables \cite{luders1950zustandsanderung} 
 is expressed by the fact that $\hat H$ 
 depends on S only through the projectors $\hat \pi_i$. Accordingly, the {\it coupling between} S {\it and} A {\it must have the form}\footnote{This form of interaction 
 can allow to describe not only ideal measurements involving well separated eigenvalues $s_i$ of $\hat s$, but also more general measurements for which the 
 projectors $\hat \pi_i$, still associated through $\hat h_i$ with  the pointer indications $A_i$, are no longer in one-to-one correspondence with the eigenvalues 
 of $\hat s$. For instance, if some $\hat \pi_i$ encompasses the eigenspaces of several different neighbouring eigenvalues, selecting the outcome $A_i$ will 
 not discriminate them, and the final state $\hat r_i=\hat\pi_i \hat r(0) \hat\pi_i / p_i$ of S will not be associated with a single eigenvalue of $\hat s$ as in an i
 deal measurement. As another example, consider two orthogonal rank-one projectors $\hat\pi_1$ and $\hat\pi_2$, coupled with sources $\hat h_1$ and 
 $\hat h_2$ that produce different outcomes $A_1$ and $A_2$, and assume that $\hat \pi_1 + \hat\pi_2$ spans the two-dimensional eigenspace associated 
 with a degenerate eigenvalue of $\hat s$;  reading the outcome $A_1$ (or $A_2$) then provides more information than this eigenvalue.} 
 $\hat H_{\rm SA} =\sum_i \hat\pi_i \otimes \hat h_i$, where $\hat h_i$ are operators of A. This form will ensure that the ``preferred basis'' is indeed the eigenbasis 
 of the projectors $\hat\pi_i$. Moreover, if $\hat s$ takes the value $s_i$, that is, $\hat\pi_i$ the value 1, the apparatus A should end up in its stable state 
 $\scriptR_i$, the pointer variable being close to $A_i$ and $\hat\Pi_i$ also taking the value 1. This  can be achieved if each $\hat h_i$ behaves as a source 
 that energetically favours relaxation towards $\scriptR_i$, thus breaking explicitly the equivalence between the various possible outcomes $A_i$. 
 (In case the pointer variable $A_i$ is an order parameter, the invariance is explicitly broken by $\hat H_\SA$.)

Likewise, $\hat H_{\rm S}$ must reduce to a linear combination of projectors $\hat\pi_i$, which only produces trivial phase factors.
 
\subsection{Assignment of a state to an ensemble of systems}\label{subsec3.2}

The analysis of the measurement process requires the assignment of a density operator to the initial state of A and the recognition of the nature of the final 
 states of S+A for the ensemble $\scriptE$ of runs and for its sub-ensembles. To this aim one may rely on the following 
 {\it maximum von Neumann entropy criterion} \cite{jaynes1957information1,jaynes1957information2,balian2006microphysics}.

\vspace{3mm}

{\it Interpretative principle 3}. Among the states compatible with the data available on an ensemble of systems, the least biased predictions are afforded by 
 assigning to it the density operator which maximises the von Neumann entropy $S(\scriptD) = - {\rm tr}\, \scriptD \ln \scriptD$.

\vspace{3mm}

This maximum entropy criterion is most often regarded as a postulate, issued from the interpretation of von Neumann's entropy as a measure of the information 
missing when only $\scriptD$  is known. However, it can be directly derived (Appendix A and Ref. \cite{ balian1987equiprobability}) from the intuitive {\it indifference or 
equiprobability principle} (that Laplace introduced under the name of principle of insufficient reason), by relying on the identification between q-expectation 
values of observables and macroscopic values of the corresponding average observables (Subsec. 2.2). 

The data are implemented in the form of constraints on the q-expectation values of some observables $\hat O_p$ ($1\le p\le p_{\rm max}$). As usual, introduction 
of Lagrange multipliers provides for the maximum entropy state $\scriptD$ a Boltzmann -- Gibbs expression, namely, the exponential of a linear combination of 
the operators $\hat O_p$. For the apparatus alone, if the equivalence between pointer values is {\it explicitly broken} by adding to the Hamiltonian $\hat H_{\rm A}$ 
the source term $\hat h_i$, and if the only constraint is about the macroscopic energy $\langle\hat H_{\rm A}+\hat h_i\rangle$, the criterion produces the canonical 
equilibrium density operator $\scriptR_i  ^\br\propto \exp[{-\beta(\hat H_{\rm A}+\hat h_i)}]$. If the Hamiltonian reduces to $\hat H_{\rm A}$, a second constraint, 
fixing the macroscopic value $A_i$ of the pointer, should be introduced to determine the density operator $\scriptR_i$ which occurs in the expected final state (1). 
A second constraint is also needed to write the expression of the initial metastable state $\scriptR(0)$
(fixing the pointer value at $m=0$ for the CW model).

\subsection{System plus apparatus in thermodynamic equilibrium}\label{subsec3.3}

Before analysing the dynamics of the measurement process (Secs. \ref{sec4} and \ref {sec5}), we determine for orientation the general form $\scriptD_{\rm eq}$ 
of the possible thermodynamic equilibrium states associated with the Hamiltonian $\hat H$ of S+A. Thermodynamic equilibrium is characterised by fixing the values 
of {\it all the conserved quantities}. Besides the constraint on energy, we must account here for the other constants of the motion, to wit, the q-expectation values 
of all the observables of S that commute with the projectors $\hat\pi_i$.
Apart from  $\beta$, the additional  Lagrange multipliers are coefficients which multiply the latter observables. 

Any equilibrium state $\scriptD_{\rm eq}$ of S+A has therefore a generalised Gibbsian form, with an exponent containing (apart from $-\beta \hat H$) an arbitrary 
operator that commutes with all the $\hat\pi_i$. Including the Lagrange multipliers, such an operator can be written as a sum $\sum_i \hat y_i$, 
where $\hat y_i$ is any operator of S acting inside the diagonal block $i$ (so that $\hat y_i=\hat \pi_i \hat y_i \hat\pi_i$).
We find therefore $\scriptD_\eq \propto  \exp(- \beta \hat H +\sum_i \hat y_i)$. Noting now that 
the full exponent, which commutes with the projections $\hat \pi_i$, has a block diagonal structure in a basis where $\hat s$ is diagonal, we can rewrite 
$\scriptD_\eq$ by exhibiting its related block diagonal structure. Finally, after separation of the various terms of 
$\hat H=\hat H_{\rm S}+\hat H_{\rm A}+\sum_i \hat\pi_i \otimes \hat h_i$, we obtain for the thermodynamic equilibrium states of S+A the general expression 

\BEQ
 \scriptD_{\rm eq}  = \sum_i q_i \hat x_i \otimes \scriptR_i ^\br,  \qquad     \sum_i q_i =1. 
  \label{eq}
\EEQ 
Each factor $q_i \hat x_i$, which arises from $\exp( -\beta \hat H_{\rm S}+\hat y_i)$, is an arbitrary non negative block diagonal operator of S, where 
$\hat x_i=\hat\pi_i \hat x_i \hat \pi_i$,  $\tr_{\rm S}\hat x_i=1$ behaves as a density operator of S. (If the eigenvalue $s_i$ is non degenerate, $\hat x_i$ 
reduces to $\hat \pi_i$.) Each factor $\scriptR_i  ^\br\propto \exp[{-\beta(\hat H_{\rm A}+\hat h_i)}]$ in (\ref{eq}) has been interpreted in Subsec. 3.2 as a 
canonical equilibrium density operator in the space of A, the source term $\hat h_i$ arising now from $\hat H_{\rm SA}$. The equilibrium states (4) of S+A 
are thus parametrized by the temperature for A, by the coefficients $q_i$ and the matrices $\hat x_i$ for S.

Distinguishing the states $\scriptR_i  ^\br$ at the macroscopic scale requires them to be characterized by different values of the pointer, close to $A_i$, with 
small variances. The operators $\hat h_i$ should therefore be sufficiently different from one another so that
the distributions tr$_{\rm A}\scriptR_i ^h \delta(A-\hat A)$ of $\hat A$ (the spectrum of which is dense) 
have single narrow peaks, well-separated for different values of $i$. The same condition will also ensure that, at the beginning of the dynamical process, 
the apparatus moves out from its metastable state $\scriptR(0)$ towards one of the equilibrium states $\scriptR^{h_i}$ so as to ensure a proper registration. 
The value of $\hat h_i$ should also be sufficiently small so that the peak of the distribution associated with $\scriptR^{h_i}$ lies close to $A_i$. These properties are 
easy to satisfy for a macroscopic apparatus\footnote{In the CW model$^{\ref{footCW}}$,  the factors $\hat h_\down=-\hat h_\up=\sum_{n=1}^N  g \hat \sigma_z^{(n)}$ 
that occur in the coupling $\hat H_{\rm SA}$ behave as a magnetic field applied to A.  The conditions for $\hat h_i$ are satisfied if $N\gg T/g$ (which lets the 
probability of the states with $m<0$ vanish for $s_z=1$), and $g<T$  (see ref. \cite{allahverdyan2013understanding}, sect. 9.4).}. 
Thermodynamic equilibrium (\ref{eq}) thus entails a complete correlation between the 
eigenvalue $s_i$ of $\hat s$ and the macroscopic value of the pointer variable.

\subsection{Measurement process as relaxation to equilibrium}\label{subsec3.4}

The states (1), (2) and (3) expected to occur after achievement of an ideal measurement process, for different ensembles, all have the equilibrium form (4) within 
replacement of $\scriptR_i  ^\br$ by $\scriptR_i $. In fact, the coupling $\hat H_{\rm SA}$ is switched off
at a time $t_{\rm decoup}$ earlier than  the end of the process. Thus, $\scriptR_i ^h$ 
can relax smoothly and reach $\scriptR_i$ at the final time $\tf$, provided the Hamiltonian $\hat H_{\rm A}$ of A does not allow direct transitions between different values 
$A_i$ (this also ensures that the states $\scriptR_i$ have a very long lifetime), and provided $\hat h_i$ is sufficiently small\footnote{In the CW model the condition 
$g<T$ ensures this relaxation (see ref. \cite{allahverdyan2013understanding}, sect. 7.2).\label{footc}} so that $\scriptR_i ^h$ lies
 in the basin of attraction of $\scriptR_i$. 
 
We have stressed (Subsec. 1.2{\it (i)}) that it is necessary (but not sufficient) to prove, by studying the dynamics of a large statistical ensemble $\scriptE$ of runs 
issued from the initial state $\scriptD(0) = \hat r(0) \otimes  \scriptR(0)$, that it ends up in the state $\scriptD(\tf)$ expressed by (\ref{fin}). We can identify (\ref{fin}) 
with a generalised thermodynamic equilibrium state (\ref{eq}), for which $\scriptR_i ^\br$ has evolved towards $\scriptR_i$ after switching off  $\hat H_{\rm SA}$.  
The free parameters of $\scriptD_\eq$ are determined from the initial condition $\scriptD(0)$, since the dynamics keeps track of the conserved quantities; 
through the identification $q_i \hat x_i=\hat\pi_i \hat r(0) \hat\pi_i\,\equiv p_i\hat r_i$ we get $q_i=p_i$ and $\hat x_i=\hat r_i$.   

We also need to prove a stronger result, still necessary and not sufficient (Subsec. 1.2{\it (ii)}). For a subset $\scriptE_\sub^{(k)}$ having yielded a proportion 
$q_i^{(k)}$ of runs with outcomes $A_i$, the corresponding final state $\scriptD_\sub^{(k)}$ should have the form (3). This final state is again recognised as 
a generalised thermodynamic equilibrium state (4), with $q_i=q_i^{(k)}$, $\hat x_i=\hat r_i$. (The property $q_i=1$, $\hat x_i=\hat r_i$ characterises
 the specific sub-ensemble $\scriptE^{(k)}_\sub = \scriptE_i$).

Thus, an {\it ideal measurement process} appears as a mere {\it relaxation of} S+A {\it to generalised thermodynamic equilibrium}, for the full ensemble $\scriptE$ 
of runs as well as for all$^{\ref{newfn2}}$ its sub-ensembles  $\scriptE_\sub^{(k)}$. In quantum mechanics, relaxation of $\scriptD(t)$ and $\scriptD_\sub^{(k)}(t)$ 
towards Gibbsian generalised thermodynamic equilibrium states (\ref{fin}) and (\ref{fin-sub}) is not granted \cite{bell1975wavepacket}. For a complete theory of ideal measurement 
processes, we must therefore justify these properties within the quantum statistical {\it dynamics} framework. We sketch the main steps of such a technical proof 
in Secs. \ref{sec4} and \ref{sec5}, as a prerequisite to the required consideration of individual runs.
 
If however one admits, in a thermodynamic scope, that the state of S+A relaxes at the final time to the equilibrium forms (2) for the ensemble $\scriptE$ and (3) 
for its sub-ensembles $\scriptE_\sub^{(k)}$, one may jump to Sec. \ref{sec6} where introduction of a last, minimalist interpretative principle will allow us to draw, 
from the expressions (2) and (3), the desired conclusions about individual measurements.

\section{Dynamics of system and apparatus for the full set of runs}
 \label{sec4}

As indicated in Sec. 1.2, the first step in the analysis of an ideal measurement process 
consists in deriving the form (2) for the final state $\scriptD(\tf)$ of S+A associated with the ensemble ${\scriptE}$, by solving the dynamical equations with 
the initial condition $\scriptD(0)=\hat r(0)\otimes \scriptR(0)$. Initiated long ago on a model \cite{daneri1962quantum}, such a task has been achieved for many other specific 
models \cite{allahverdyan2013understanding}. We only survey here the formal features of the solution in the general case, postponing any interpretation.

Since the measurement problem is related to the foundations of physics, a theoretical analysis should rely on the most fundamental dynamical law, that is, 
the Liouville--von Neumann equation $i\hbar \d \scriptD(t)/\d t=[\hat H, \scriptD(t)]$ which governs an isolated, large but finite system. It is therefore preferable 
(but not compulsory) to consider that A includes the needed thermal bath or environment so that S+A is {\it isolated}. Taking then into account the  above form 
of $\hat H$ including the interaction $\hat H_{\rm SA} =\sum_i \hat\pi_i \otimes \hat h_i$, and the approximate commutation $[\hat H_{\rm S}, \hat r(0)] \simeq 0$ 
which ensures that the marginal state $\hat r(t)$ of S is perturbed only by the interaction $\hat H_{\rm SA}$  during the process, we check that  $\scriptD(t)$ can 
be parameterised as 

\BEQ
\scriptD(t) = \sum_{i, j} \hat\pi_i \hat r(0) \hat\pi_j \otimes\scriptR_{ij}(t)  
\label{Dtsum}
\EEQ
 in terms of a set  $\scriptR_{ij}(t)=\scriptR_{ji}^\dagger (t)$  of operators in the Hilbert space of A. 
 The latter operators must be found by solving  the equations of motion

 \BEQ 
 \hspace{-3mm}
 i \hbar \frac{\d \scriptR_{ij}(t) }{\d t} = (\hat H_{\rm A} +\hat h_i) \scriptR_{ij}(t) - \scriptR_{ij}(t) (\hat H_{\rm A} +\hat h_j), 
 \label{dRdt} \EEQ
with the initial conditions  $\scriptR_{ij}(0)=\scriptR(0)$. 

The dynamics thus involves {\it solely the apparatus. }
Its coupling with the tested system occurs in (6) only through $\hat h_i$ and $\hat h_j$, a specific 
property of ideal measurements. 
 Ideality involves separation of S from A upon achievement of the measurement, meaning that each $\hat h_i$ is switched off at 
the last stage of the process. Moreover, {\it the dynamics of each block} $\scriptR_{ij}$ of the density matrix $\scriptD$, whether $i=j$ or $i>j$,  {\it is decoupled 
from the dynamics of the other blocks}. (The  $i<j$ blocks follow by Hermiticity.)

If the environment is regarded as external to the apparatus, with weak interactions, its elimination from the equations of motion, 
achieved by standard methods of quantum statistical mechanics, produces additional terms in (6). However the decoupling still takes place.

In any case, the evolution of $\scriptD(t)$ towards the equilibrium state $\scriptD(t_f)$ is an irreversible process, during which the coarse grained entropy increases. 
The compatibility of this feature with the reversibility of the differential equations (6) is ensured by the principle 2 of Subsec. 2.3, which allows us to disregard 
physically irrelevant elements issued from the exact equations, and thus to justify approximations of quantum statistical mechanics. The macroscopic number of 
degrees of freedom for the bath and for the pointer included in A, and a suitable choice of parameters in $\hat H_{\rm A}$ and $\hat H_{\rm SA}$ will therefore be 
needed, for each model, to explain the required relaxations and to estimate their time scales, as illustrated by the CW model$^{\ref{footCW}}$. In decoherence 
approaches, which focus on the disappearance of the off-diagonal blocks $\scriptR_{ij}$ for $i\neq j$, irreversibility is ensured by the large size of an external 
environment \cite{zurek2003decoherence,laloe2012we}.

Two types of relaxation, with different time scales, arise independently from the dynamical equations (\ref{dRdt})\footnote{Authors do not always give the same 
meaning to the various words used. We term as {\it truncation} the disappearance of the off-diagonal blocks of the density matrix of S+A 
under the effect of an arbitrary mechanism (including dephasing),  and specialise {\it decoherence} to the production of this effect by interaction with 
an environment or a thermal bath.  We term as {\it registration} the process which leads each 
diagonal block to the correlated state $\hat r_i\otimes \scriptR_i$, and as {\it reduction} 
the transition from $\hat r(0)$ to some $\hat r_i$  for an individual run.\label{meanings}}.

 ($i$) {\it ``Truncation'':}  For $i\neq j$, the coherent contributions $\scriptR_{ij}(t)$ decay for all practical purposes owing to the difference between 
$\hat h_i$ and $\hat h_j$, and rather quickly vanish. The off-diagonal blocks of the density matrix $\scriptD(t)$ are thus truncated as regards the physically 
attainable observables\footnote{The matrix elements of  $\scriptR_{ij}(t)$ with $i\neq j$ contain rapidly oscillating phase factors. As for any irreversible process, 
physical quantities involve sums over very many of them, which cancel out for times less than the huge recurrence time. So for all practical purposes they can be 
omitted  after the relaxation time owing to the macroscopic size of the apparatus, in spite of the constant value of the sum 
 tr$_{\rm A} \scriptR_{ij}(t) \scriptR^\dagger_{ij}(t)$ of the modulus square of the matrix elements of $\scriptR_{ij}(t)$. However, would one wish to calculate 
mathematical objects, for instance to check that the exact von Neumann entropy (without coarse graining) remains constant, they would definitely contribute.\label{fn1212}}.
Depending on the model, this decay may be governed by different mechanisms\footnote{This ``truncation'' process has abundantly been studied in the literature on 
measurements. It is often supposed to be the result of a decoherence produced by a coupling with an external environment. In the case of measurements, the 
off-diagonal blocks to be suppressed by the dynamics (6) are those which relate different eigenvalues $s_i$ of $\hat s$ in a basis diagonalizing $\hat s$. 
However, as discussed in Subsec. 3.1, S must be coupled to A (including the environment) by an interaction of the form 
$\hat H_{\rm SA}=\sum_i \hat\pi_i \otimes\hat h_i$, where each operator $\hat h_i$ should ensure relaxation towards the equilibrium state $\scriptR_i^h$ of A. 
Thus, explaining the truncation by a decoherence process may be satisfactory only if the coupling with an external environment has a particular form depending 
both on the tested system and on the pointer observable, so that S+A is piloted by a potential $\hat H_{\rm SA}$ of the above type 
(see ref. \cite{allahverdyan2013understanding}, sect. 2.7). 
In the CW model$^{\ref{footCW}}$, several mechanisms occur, involving or not a thermal bath. Over the short time scale $\hbar/g \sqrt{N}$, truncation results 
(see ref. \cite{allahverdyan2013understanding}, sect. 5) from the dephasing$^{\ref{fn1212}}$ between 
the oscillations yielded by the factor $\exp 2i t\hbar^{-1}$$ \sum_{n=1}^N g\hat \sigma_z^{(n)} $ 
entering $\scriptR_{\up \down}(t)$, which have different frequencies (due to the randomness of $\sigma_z^{(n)}$ in the initial paramagnetic state of A). 
Information is thereby lost through a cascade of correlations of higher and higher order, less and less accessible, between $\hat s_x$ or $\hat s_y$ and
the spins of A,  in such a way that $\scriptR_{\up \down}(t)$ practically tends to zero as regards the accessible observables. Recurrences are wiped out 
(see ref. \cite{allahverdyan2013understanding}, sect. 6), 
either by the coupling $\gamma$ with the phonon bath (provided $T/J\gg \gamma\gg g/NT$), or by a spread $\delta g$ in the couplings $g$ of $\hat H_{\rm SA}$  
(provided $\delta g\gg g/\sqrt{N}$).}.

($ii$) {\it ``Registration'':} For $i=j$, the evolution of $\scriptR_{ii}(t)$ governed by (\ref{dRdt}) is a mere relaxation from the metastable state $\scriptR(0)$ to the 
equilibrium state $\scriptR_i^h$ in the presence of the source $\hat h_i$, and then to $\scriptR_i$ after $\hat H_{\rm SA}$ is switched off. The correlation 
between $s_i$ and $A_i$ needed to register the outcome is thereby established\footnote{While much attention has been paid 
to the vanishing of the off-diagonal blocks, the relaxation of the diagonal blocks is too often disregarded, although it produces the correlations that ensure the 
possibility of reading the outcome. In the CW model (see ref. \cite{allahverdyan2013understanding}, sect. 7), 
this process is triggered by $\hat h_i$ which makes $\scriptR(0)$ unstable and should 
be sufficiently large to exclude false registrations ($g\gg J/\sqrt{N}$). Later on, the relaxation of $\scriptR_{ii}(t)$ to $\scriptR_i^h$, and finally to $\scriptR_i$ 
after $\hat H_{\rm SA}$ is switched off, is governed by the dumping of free energy from the magnet to the phonon bath; its characteristic duration is the 
registration time $\hbar/\gamma(J-T)$.}. Since registration requires a dumping of free energy into the bath, it its typically slower than truncation.

These two irreversible processes are unrelated and should not be confused. The registration consists in the {\it establishment of correlations between the pointer 
and the tested observable}, that we generally denoted as $\hat s$ (for the CW model, this tested variable is the component $\hat s_z$ of the spin), whereas 
the truncation proceeds through
gradual creation and subsequent {\it vanishing of correlations between the pointer and observables that do not commute with} $\hat s$ (for the CW model, these 
observables are the transverse components $\hat s_x$ and $\hat s_y$). Both are essential\footnote{In an analogy to Nuclear Magnetic Resonance, truncation is 
similar to ${\cal T}_2$ processes and generally much faster than registration, which bears some analogy to ${\cal T}_1$ processes.} .

Thus, microscopic dynamics confirms the surmise of relaxation towards the generalised thermodynamic equilibrium state (\ref{fin}) for S+A in the ensemble 
$\scriptE$.  As S and A have been decoupled at some time $t_{\rm decoup}$ before $\tf $, the remainder of our discussion  will involve {\it only the apparatus}.

\section{Through sub-ensembles towards individual runs} \label{sec5}

The expression $\scriptD(\tf)=\sum_i p_i \scriptD_i$ of the final state of S+A thus derived for the large ensemble $\scriptE$ of runs might suggest that the task is over. 
It seems to mean that the set $\scriptE$ gathers, as expected, a proportion $p_i$ of individual runs having ended up in the state $\scriptD_i$. However, as already 
indicated in Subsec. 1.2 ({\it i}) and explained in detail below (Subsec. 5.1), Schr\"odinger's quantum ambiguity 
makes it {\it fallacious to postulate directly} such an interpretation  of the separate terms $p_i \scriptD_i$ of $\scriptD(\tf)$. In order to justify it, 
we adopt the following strategy. We will first draw (Sec. 5) further information from the dynamics of sub-ensembles in a formal quantum frame, and this will allow 
us to introduce afterwards an indisputable interpretative principle (Sec. 6).

\subsection{Quantum ambiguity of mixed states} \label{subsec5.1}

Classical probabilities presuppose the existence of a sample space, so that an ordinary probability distribution can be identified with the set of relative frequencies 
of occurrence of some property among a large number of individual events. The construction of classical sub-ensembles relies on the 
possibility of distinguishing individual events so as to select part of them$^{\ref{newfn2}}$. One can thus readily infer from a classical probability distribution 
a {\it unique set of sub-ensembles} $\scriptE^{(k)}_\sub$ and their associated distributions $\scriptD^{(k)}_\sub$.

The situation is different in quantum mechanics. There also, the assignment of a density operator to an ensemble of systems is a means of making statements about 
experimental facts pertaining to this ensemble, but experiments performed with different apparatuses can provide results (such as the violation of Bell's inequalities) 
which are not compatible with the existence of a sample space describing individual systems.

Formally, this difficulty is expressed by Schr\"odinger's {\it quantum ambiguity of the decompositions} of a density operator \cite{schrodinger1935discussion,schrodinger1936probability,park1988thermodynamic}, 
which we illustrate by the simple, well known example of an unpolarised ensemble $\scriptE$ of spins in the state $\half\hat I$. The decomposition 
 $\hat I= | z\rangle\langle z| \,  +  | \!\! -\! z\rangle\langle -z|$ for this state seems to mean that half of the spins are polarised along 
 $| \hspace{-1mm}+\! z\rangle$, the other half along $| \hspace{-1mm}-\! z\rangle$. 
 However, the same argument applied to the alternative decomposition  $\hat I=| x\rangle\langle x| \, + \,| \!\! -\! x\rangle\langle -x|$ would imply 
 that $\scriptE$ might be split into four sub-ensembles, each of which would gather spins polarised simultaneously in two orthogonal directions, which is nonsensical. 
 And there exist many other decompositions, suggesting interpretations contradictory to each other, hence meaningless.
 Likewise, it is obviously inconsistent to interpret the state of an unpolarised spin extracted 
 from a singlet pair as a mixture of completely polarised spins\footnote {One could argue that an interpretation might arise from the knowledge of the preparation 
 of the ensemble $\scriptE$. If $\scriptE$ is built by putting together two equal-sized sub-ensembles of spins polarised in the directions $| z\rangle$ and  
 $|\hspace{-1mm} -\! z\rangle$, respectively, and provided we keep track of the origin of each sample, it is legitimate to interpret separately each term of 
 $\half\hat I=\half | z\rangle\langle z| \, + \half | \!\! -\! z\rangle\langle -z|$. However, if the two sub-ensembles have merged at random, only $\half\hat I$ is meaningful. 
 No experiment can allow to distinguish two different preparations of the ensemble $\scriptE$ having led to the same mixed state, or to distinguish different 
 populations within $\scriptE$, if no other information than this state is available.}. 

The above argument is general. Any mixed state $\scriptD$ can be decomposed in an infinity of ways as a weighted sum of projectors onto pure states, 
 which need not be orthogonal and which cannot be decomposed further\footnote {All other decompositions, involving mixed states, are built by grouping terms 
 of the decompositions in terms of pure states}. Attempting to interpret simultaneously two different decompositions would lead to contradictions. 
 Due to this quantum ambiguity, the irreducible nature of q-probabilities forbids the recognition of a sample space that would refer to individual systems 
 and would underlie density operators. Hence, if nothing else than $\scriptD$ is known, the {\it logical incompatibility between arbitrary mathematical decompositions 
 prevents us from giving a physical meaning} to the separate terms of such a decomposition.
 
For measurements, once the expression (2) of $\scriptD (\tf)$ has been globally derived as in Sec. 4, the existence of mathematical decompositions of 
 $\scriptD (\tf)$ incompatible with the particular one $\scriptD (\tf ) = \sum_i p_i\scriptD_i$ makes it unjustified to 
 bluntly infer (as is often done)  that each individual run ends up in one 
 or another of the states $\scriptD_i$. There, extra information will be searched by noting that the runs are expected to be tagged {\it after measurement} by the 
 indication of the pointer, allowing the consideration of sub-ensembles. The idea that the same dynamical equations govern both $\scriptE$ and its sub-ensembles 
 will help us to pursue within the quantum formalism as far as possible, postponing interpretation so as to introduce weakest possible interpretative principles. 
 Our next task (Subsec. 5.4) will therefore consist in proving that, for {\it all possible physical sub-ensembles}$^{\ref{newfn2}}$
 $\scriptE_\sub^{(k)}$ of  $\scriptE$, S+A  ends up in a state of the form (\ref{fin-sub}),  
 $\scriptD_\sub^{(k)} = \sum_i q_i^{(k)} \scriptD_i$. 

\subsection{Introduction of sub-ensembles} \label{subsec5.2}

We remind (Subsec. \ref{subsec2.4}) that a given individual system can statistically be described by different quantum states, depending on our information 
about the physical sub-ensemble in which it is embedded. These states, usually mixed, are related to one another. 
When two disjoint sub-ensembles, $\scriptE_\sub^{(k)}$ (with $\scriptN_\sub^{(k)}$ elements) and $\scriptE_\sub^{(k')}$ (with $\scriptN_\sub^{(k')} $ elements), 
described by $\scriptD_\sub^{(k)}$ and $\scriptD_\sub^{(k')}$, respectively, merge to constitute an ensemble $\scriptE$ described by $\scriptD$ 
(with $\scriptN=\scriptN_\sub^{(k)}+ \scriptN_\sub^{(k')}$ elements), 
the q-expectation values defined by the correspondence  $\hat O\mapsto \langle\hat O\rangle$ for $\scriptE$, $\scriptE_\sub^{(k)}$ and $\scriptE_\sub^{(k')}$ 
have the same additivity property as ordinary averages. This is expressed at each time by\footnote{If an ensemble $\scriptE$ 
is constructed by putting together two ensembles $\scriptE_\sub^{(k)}$ and $\scriptE_\sub^{(k')}$, the ingredients $\lambda$, $\scriptD_\sub^{(k)}$ and 
$\scriptD_\sub^{(k')}$ of the decomposition (7) of the state $\scriptD$ associated with $\scriptE$ keep a physical meaning as long as the sub-ensembles 
$\scriptE_\sub^{(k)}$ and $\scriptE_\sub^{(k')}$ can be identified within $\scriptE$. However, if track is lost of these sub-ensembles within $\scriptE$, 
the two terms in $\scriptD$ cannot be determined from any experiment performed by extracting samples at random from $\scriptE$.}

\BEQ \label{Dksplit}
\scriptD(t)=\lambda \scriptD_\sub^{(k)}(t)+(1 -\lambda)\scriptD_{\sub}^{(k')}(t), 
\EEQ
with the weight $\lambda=\scriptN_\sub^{(k)}/\scriptN$. All three states $\scriptD(t)$, $\scriptD^{(k)}_\sub (t)$ and $\scriptD^{(k')}_\sub (t)$ are governed by the same 
dynamical equations, involving the Hamiltonian that characterises the considered system.

However, conversely, due to the matrix nature of $\scriptD$, there exist many operators $\scriptD_\dec$ issued from decompositions of the type (7) 
which cannot be associated with sub-ensembles and have no physical meaning. 
In fact, the mathematical decompositions of $\scriptD$ depend on continuous parameters, so that the number 
of states $\scriptD_\dec$ is infinite. In contrast, if the ensemble $\scriptE$ has $\scriptN$ elements, the number $2^\scriptN-1$ of its subsets (containing at least 
one element) is finite. For large $\scriptN$, the number of physical sub-ensembles $\scriptE_\sub^{(k)}$ described by states $\scriptD^{(k)}_\sub$ is still smaller, 
growing polynomially with $\scriptN$, because each $\scriptE_\sub^{(k)}$ should contain many elements. Thus, only a tiny proportion of decompositions 
of the type (7) may describe a splitting of $\scriptE$ into physical sub-ensembles. 

The impossibility of extracting from $\scriptD$ alone information about individual runs and even about sub-ensembles is the form taken here by the 
{\it measurement problem}.
In order to overcome it, we need a criterion allowing a quantum description of physical sub-ensembles of runs extracted from  $\scriptE$. {\it After achievement} of the 
measurement process, observation and selection of the pointer indications would afford identification of the sub-ensembles $\scriptE_\sub^{(k)}$ characterised by the 
proportions $q_i^{(k)}$ of runs that have produced the outcome $A_i$. We expect a state $\scriptD_\sub^{(k)}(\tf)$ of the form (3) to be assigned to each one, 
and we also expect the same state to describe the sub-ensemble at the final time {\it just before reading}. It is then natural to postulate that, at least {\it during 
the very last stage} $\tpf <  t < \tf$ of the process, one can associate quantum states $\scriptD_\sub^{(k)}(t)$ to the sub-ensembles $\scriptE_\sub^{(k)}$ although 
the latter cannot yet be identified.  Hence we state:

\vspace{3mm}

{\it Interpretative principle 4}. Density operators which obey the probabilistic and dynamic rules of quantum mechanics may be assigned 
not only to a large statistical ensemble of systems, but also to any one of its physical sub-ensembles. Such a simultaneous assignment of {\it several 
sub-ensemble dependent states} to similar systems can be done during a short delay preceding the time when the sub-ensembles will be 
identified through some macroscopic property.

\vspace{3mm}

This principle implies that the evolution of the density operators $\scriptD_\sub^{(k)}(t)$ 
during the time lapse $\tpf <  t < \tf$ is governed by the same equations as 
for $\scriptD(t)$. (This is consistent with the fact that, in the Heisenberg picture, the dynamical equations do not depend on the state.) We wish to work out these 
dynamical equations from $\tpf$ to $\tf$ so as to prove that $\scriptD_\sub^{(k)}(t)$ relaxes to the expected form (3). However we have a priori no information 
about $\scriptD_\sub^{(k)}(\tpf)$ at the new initial time $\tpf$, except for the fact that $\scriptD_\sub^{(k)}(\tpf)$ is an element of some 
decomposition (7) of $\scriptD(\tpf)$, a property which will yield constraints on this initial state. Note that, if the state $\scriptD_\sub^{(k)}$ entering (7) 
were associated with a sub-ensemble picked at random from $\scriptE$, it would be for large $\scriptE$ the same as $\scriptD$ itself. 
The sub-ensembles $\scriptE_\sub^{(k)}$ of interest are therefore scarce$^{\ref{newfn2}}$ within $\scriptE$.  
 
Individual runs and sub-ensembles evidently exist experimentally at all times; they can be tagged and followed during the whole process
 (``waiting for the outcome''). Nevertheless, we 
cannot consider theoretically the physical sub-ensembles $\scriptE_\sub^{(k)}$ at arbitrary times. We can acquire information about the initial state 
$\scriptD^{(k)}_\sub(\tpf)$ only from Eq. (7), through knowledge previously obtained about the state $\scriptD(\tpf)$ associated with the full ensemble $\scriptE$. 
We therefore choose $\tpf$ {\it sufficiently late}\footnote{At earlier times, the form of $\scriptD(t)$ would provide weaker constraints on the initial state of 
$\scriptD^{(k)}_\sub(t)$. Still earlier and especially at the beginning of the process, no reasonable splitting of $\scriptD(t)$ even exists. The very possibility of 
considering the quantum states $\scriptD^{(k)}_\sub(t)$ at the time $\tpf$ and hence later emerges from the relaxation of the state $\scriptD(t)$ which occurred 
at earlier times $t$.} so that the interaction $\hat H_\SA$ between A and S has been switched off and that this state $\scriptD(\tpf)$ has {\it already reached 
the final form} (2). We also take $\tpf$ {\it sufficiently early} so that the relaxation time for each sub-ensemble is shorter than the duration $\tf-\tpf$ of the 
evolution (Subsec. \ref{subsec5.4}). 

\subsection{Ingredients of the dynamics} \label{subsec5.3}

We wish to prove that {\it all the states} $\scriptD_\sub^{(k)}(t)$ associated with physical subsets of runs {\it end up in the required form} 
$\scriptD_\sub^{(k)}(\tf )=\sum_i q_i^{(k)} \scriptD_i$. This property might be regarded as intuitive, since this is just a relaxation towards a generalised 
{\it thermodynamic equilibrium} state, often supposed to be ensured by an environment. However, even though the probability distribution $\scriptD(\tpf)$ 
associated with the {\it full ensemble} $\scriptE$ has already reached its equilibrium form $\scriptD(\tf) = \sum_i p_i \scriptD_i$, the distributions 
$\scriptD_\sub^{(k)}(\tf )$ associated with {\it its sub-ensembles} may still be off equilibrium. A {\it dynamical derivation} is necessary to establish their relaxation rigorously.

Since S and A have been decoupled at the time $t_{\rm decoup}$ before $\tpf$, the Hamiltonian reduces for $t>\tpf$ to $\hat H_{\rm S}+\hat H_{\rm A}$. 
To simplify the discussion, we assume here that the eigenvalues of $\hat s$ are non degenerate, hence  $\hat r_i=\hat\pi_i=|s_i\rangle\langle s_i|$, 
and that $\hat H_{\rm S}=0 $\footnote{For degenerate eigenvalues $s_i$, the only change in the forthcoming derivation, if the states 
 $\hat r_i \equiv |i\rangle\langle i|$ are pure, is the replacement of  $ |s_i\rangle$ by the ket $|i\rangle$ in the eigenspace of $\hat s$ associated with $s_i$. 
 If the density operator $\hat r_i$ is mixed, we note that this operator of S is not modified by the process, while remaining fully coupled with $A_i$ for $t>\tsplit$. 
 We should therefore preserve this property when we consider the decompositions (7) of $\scriptD$ which produce the states $\scriptD_\sub^{(k)}$ of physical 
 sub-ensembles $\scriptE_\sub^{(k)}$. The poly-microcanonical relaxation of A then produces again the final state (10). A non-vanishing $\hat H_{\rm S}$ 
 would generate for each $i$ a different phase factor, which is ineffective.}. 
 The dynamics of $\scriptD_\sub^{(k)}(t)$ is therefore governed for $t>\tpf$ by the Hamiltonian $\hat H_{\rm A}$ {\it of the apparatus alone}.

We now need to characterise the initial states $\scriptD_\sub^{(k)}(\tpf)$. These operators cannot be fully determined, but they must arise from some decomposition 
 (7) of $\scriptD(\tpf) = \scriptD(\tf) = \sum_i p_i \scriptD_i = \sum_i p_i \hat\pi_i \otimes \scriptR_i$, a density operator that we first analyse. The state $\scriptR_i$ 
 describes canonical equilibrium of the apparatus, with moreover a constraint on the macroscopic value $A_i$ for the pointer (Subsec 3.3). As A is macroscopic, 
 the fluctuations of $\hat H_{\rm A}$ around $\langle \hat H_{\rm A}\rangle$ and of the pointer observable $\hat A$ around $A_i$ are small in relative size, and it 
 is legitimate to replace in $\scriptD(\tpf)$ the {\it canonical} equilibrium states $\scriptR_i$ of A by {\it microcanonical} ones, $\scriptR_i^\mu$, as regards both the 
 energy and the pointer variable. Thus, within the Hilbert space of A, we denote as $|A_i, \eta\rangle$ a basis of kets constrained by the fact that the macroscopic 
 energy lies in some small range and that $\hat A$ also lies between $A_i -\Delta$ and $A_i + \Delta$, where $2\Delta$ is larger than the width of $\scriptR_i$ 
 (Sec. 1.1).  As the spectrum is dense, the index $\eta$ may take a very large number $G_i$ of values. We have denoted as $\hat\Pi_i$ the 
 projector$^{\ref{newfootone}}$ over the eigenspace of $\hat A$ associated with the eigenvalues lying between $A_i -\Delta$ and $A_i + \Delta$ 
 (for arbitrary energies), hence, $\scriptR_i^\mu  \hat\Pi_j = \scriptR_i^\mu \delta_{ij}$. The equivalence between the canonical and microcanonical states 
 $\scriptR_i$ and $\scriptR_i^\mu$ is expresed by $\tr_{\rm A} \scriptR_i \hat\Pi_j \simeq {\rm tr}_{\rm A} \scriptR_i^\mu   \hat\Pi_j = \delta_{ij} $\footnote{Strictly 
 speaking, this equality holds only for the microcanonical states $\scriptR_i^\mu$. However, $\tr_{\rm A} \scriptR_i \hat\Pi_i$ is close to 1 if $\Delta$ is sufficiently
  large compared to the width of $\scriptR_i$, and tr$_{\rm A} \scriptR_i \hat\Pi_j$ for $i\neq j$ is negligible if $\Delta$ is small compared to the distance between 
  the possible outcomes $A_i$.\label{fnnoot}}. The microcanonical equilibrium state of A is then proportional to a projector:
 
 \BEQ		
 \hspace{-0cm}
 \scriptR_i^\mu = \frac{1}{G_i} \sum_\eta |A_i, \eta \rangle\langle A_i, \eta |.
\label{micro}
\EEQ

Accordingly, the state $\scriptD(\tpf) \simeq \sum_i p_i \hat\pi_i \otimes \scriptR_i^\mu$ (where $\hat\pi_i=|s_i\rangle\langle s_i|$) does not lie in the full Hilbert space 
$\scriptH$ of S+A, but in its small, {\it shrunken subspace} $\scriptH_\shr$ spanned by the kets $|s_i\rangle |A_i, \eta\rangle$. In this subspace, the tested system 
and the pointer value are {\it correlated}. Since the initial state $\scriptD_\sub^{(k)}(\tpf)$ associated with the sub-ensemble $\scriptE_\sub^{(k)}$ must be an 
element of some decomposition (7) of $\scriptD(\tpf)$, it is also constrained to {\it lie in the subspace} $\scriptH_\shr$. It must therefore have the form
 
\BEQ
\hspace{-6mm} \label{Dsubdec}
\scriptD_\sub^{(k)}(\tsplit)=\sum_{i, j, \eta, \eta'} |s_i\rangle |A_i, \eta\rangle  K^{(k)}(i, \eta; j, \eta';\tpf) \langle s_j| \langle A_j, \eta'|, \nn\\
\hspace{6.5cm}  (9) \nn
\EEQ 
\setcounter{equation}{9}
\hspace{-2mm}
where $K^{(k)}$ is a Hermitean, normalised and nonnegative matrix, which however remains unknown\footnote{All mathematical decompositions of $\scriptD(\tpf)$ 
having the form (7)  give rise to arbitrary operators of the form (9). According to the principle 4, some of these operators (but we cannot determine which ones) 
describe physical sub-ensembles, and we consider only these, whereas the other ones, 
much more numerous, are physically meaningless as discussed in Subsecs. 5.1 and 5.2.}.
  
\subsection{Poly-microcanonical relaxation} \label{subsec5.4}

We now consider the dynamics of $\scriptD_\sub^{(k)}(t)$ for $t > \tpf$, governed {\it by the Hamiltonian $\hat H_{\rm A}$ of the sole apparatus} and starting from 
  the partially unknown initial condition (9). As $\scriptD_\sub^{(k)}(t)$ is an element of some decomposition (7) of $\scriptD(t)$ which is constant, it remains in the 
shrunken  subspace $\scriptH_\shr$  and retains the form (9) where $K^{(k)}$ depends on time. Only the part of $\hat H_{\rm A}$ that lives in the Hilbert subspace 
  $\scriptH_\shr$ is relevant for the evaluation of the time dependence of $K^{(k)}(i, \eta; j, \eta';t)$. 
  We will rely for the sub-ensembles on a new relaxation mechanism \cite{allahverdyan2013understanding}, which we term here as  {\it ``poly-microcanonical''}. 
  One can regard it as a generalisation of the standard microcanonical relaxation \cite{vanHove1962,vanKampen1962,landau1980statistical,bander1996approach} which, for 
  any initial state in the {\it only Hilbert subspace} $\scriptH_i$ of $\scriptH$ spanned by the kets $|s_i\rangle |A_i, \eta\rangle$ for given $i$, 
  produces a decay towards the state (8) proportional to the projector on $\scriptH_i$. 

Here, we consider the subspace $\scriptH_\shr$ of $\scriptH$, the direct sum of {\it several microcanonical subspaces} $\scriptH_i$. 
We assume that weak interactions in the apparatus (including the environment) induce 
  among the kets $|A_i, \eta\rangle$ rapid transitions {\it within each subspace} $\scriptH_i$. Such interactions are realistic for a macroscopic apparatus; they have 
  little effect on the processes described in Sec. \ref{sec4}. In each elementary transition, $\eta$ is modified while both the macroscopic  energy and the macroscopic 
  pointer value are not affected. The absence of jumps between different pointer values is needed to ensure the stability of the states $\scriptR_i$. Owing to this 
  conservation of macroscopic quantities, the process is very rapid\footnote{Two different mechanisms achieving such a process have been fully worked out for 
  the CW model (see ref. \cite{allahverdyan2013understanding}, sec. 11.2), and it has been shown that they produce the result (\ref {Dsubfdec}). 
  In the more realistic one (see ref. \cite{allahverdyan2013understanding},  Appendices H and I), 
  the transitions that modify $\eta$ are produced by an interaction $\hat V$ between the magnet and the bath which has a variance $v^2={\rm tr}\,\hat V^2$; 
  an average delay $\theta$ separates successive transitions. The poly-microcanonical relaxation may take place even if $\hat V$ is not macroscopic, with a 
  variance scaling as $v\propto N^a$ ($a<1$) for large $N$. For a short $\theta$ that scales as $\theta\propto 1/N^b$ ($a<b<2a$), the characteristic time 
  $\tau_\sub=\hbar^2/v^2\theta$ scales as $1/N^c$ where $c=2a-b$, $0<c < a<1$; it is short compared to the registration time, which dominates $\tf$ because 
  registration involves a macroscopic dumping of energy from the magnet to the bath, in contrast to the poly-microcanonical relaxation. \label{footf}}. 

The poly-microcanonical relaxation is thus a ``quantum collisional process'', irreversible for a large apparatus. Acting separately in each sector, on both sides 
  $|A_i, \eta\rangle$ and $\langle A_j, \eta'|$ of (\ref{Dsubdec}), it produces two different effects. ($i$) For $i=j$, the result is the same as for the standard 
  microcanonical relaxation. All terms $\eta\neq\eta'$ disappear from $K^{(k)}(i, \eta; i, \eta';t)$, while the terms $K^{(k)}(i, \eta; i, \eta;t)$ all tend to one another, 
  their sum remaining constant. Altogether, the coherences disappear and {\it the populations equalise within each sector}. ($ii$) For $i\neq j$, all contributions 
  to (\ref{Dsubdec}) fade out and eventually vanish, so that {\it the different sectors i become uncorrelated}. Both effects occur over the {\it same time scale} 
  $\tau_\sub$, which (by definition of $\tpf$) is shorter$^{\ref{footf}}$  than $\tf-\tpf$. As the mechanism is already effective before  $\tpf$, the relaxation is likely 
  to have already been effective at $\tpf$. Anyhow, $\scriptD_\sub^{(k)}$ reaches at the final time $\tf > \tsplit+\tau_\sub$ the ``poly-microcanonical'' 
  equilibrium\footnote{The present process should not be confused with those of Sec. 4. On the one hand, in contrast to the latter, it involves only the 
  apparatus (which includes a bath or an environment). On the other hand, it requires the achievement of both the truncation and the registration.\label{newestfn}}
 
\BEQ \label{Dsubfdec}
\scriptD_\sub^{(k)}(\tf) = \sum_i q_i^{(k)} \hat r_i \otimes \scriptR_i^\mu, \quad  \nn\\
   q_i^{(k)}=\sum_\eta K^{(k)}(i, \eta; i, \eta;t'_{\rm f})=\tr \, \scriptD^{(k)}_\sub(\tpf) \, \hat\Pi_i 
\EEQ

This general expression for the final state $\scriptD^{(k)}_\sub(\tf)$ associated with any physical sub-ensemble 
depends on the initial condition (9) only through the coefficients $q_i^{(k)}$. 
The distinction between canonical and microcanonical equilibria being macroscopically irrelevant, we have derived within the quantum dynamical formalism 
the {\it relaxation to the required equilibrium form} (3) {\it for arbitrary sub-ensembles} $\scriptE_\sub^{(k)}$. 
All of these involve at the final time the {\it same building blocks} $\scriptD_i$, so that {\it the quantum ambiguity has been removed}.

\subsection{Properties of the coefficients $q_i^{(k)}$} \label{subsec5.5}

The form (3) or (10) for the final state associated with any sub-ensemble $\scriptE_\sub^{(k)}$, together with the property$^{\ref{fnnoot}}$
$\tr_{\rm A} \scriptR_i \hat\Pi_j=\delta_{ij}$, imply 
	
\BEQ	\tr\,\scriptD_\sub^{(k)}(\tf) \hat\Pi_i = q_i^{(k)}.       
\EEQ
Each weight $q_i^{(k)}$ is therefore identified as the {\it q-probability} of occurrence of the macroscopic value $A_i$  for the pointer in the sub-ensemble 
$\scriptE_\sub^{(k)}$ of runs of the measurement. The narrowness ($\Delta \ll |A_i - A_j|$) of the spectrum of the projectors $\hat\Pi_i$ entails that for any 
$\scriptE_\sub^{(k)}$ the {\it q-distribution} $\tr\,\scriptD_\sub^{(k)}(\tf) \delta(\hat A - A)$ of $\hat A$ is strongly peaked around the values $A_i$, with the 
weights $q_i^{(k)}$. These quantum properties are still formal and call for an interpretation (Sec. \ref{sec6}).

We also note, by using the commutation $[\scriptR_i, \hat\Pi_j]=0$, that the q-expectation values 

\BEQ	
\hspace{1.5cm}
\tr\,\scriptD_\sub^{(k)}(\tf) [\hat\Pi_i, \hat O] = 0      \hspace{2cm} (12{\rm a}) \nn
\EEQ
vanish for arbitrary operators $\hat O$ of S+A. More generally, if the typical dimension $G$ of the projectors $\hat \Pi_i$ is large, and if $\hat P$, $\hat P'$ and 
$\hat P''$ denote arbitrary projection operators with finite dimension whereas $G\gg 1$, one readily shows, by expansion on the basis $|s_i\rangle|A_i, \eta\rangle$, 
that

\BEQ		\hspace{0.5cm}
\tr\,\scriptD_\sub^{(k)}(\tf) \hat P'[\hat\Pi_i, \hat P]\hat P'' = {\cal O}\left(\frac{1}{G}\right)	        \hspace{1.3cm}	 (12{\rm b}) 	\nn
\EEQ  
is small. Any operator of S+A containing as a factor a commutator $[\hat\Pi_i, \hat O]$ of a pointer observable with an arbitrary observable $\hat O$ 
(finite for large $G$) can be written as a weighted sum of terms (12b). Hence, Eqs. (12a-b) express that the {\it q-expectation value, in the final state}, of  any operator 
depending on the projectors $\hat\Pi_i$ through {\it commutators $[\hat\Pi_i, \hat O]$ with arbitrary finite observables $\hat O$, is negligible} for a macroscopic pointer.
We shall rely on this property in Subsection 6.1.

Finally the coefficients $q_i^{(k)}$ that characterise the states (10) derived for the whole collection of sub-ensembles $\scriptE_\sub^{(k)}$ of $\scriptE$ possess a 
hierarchic structure embedded in the following additivity property. If some sub-ensemble $\scriptE_\sub^{(k)}$ is split into two smaller sub-ensembles 
$\scriptE^{(k')}_\sub$ and $\scriptE^{(k'')}_\sub$, containing $\scriptN^{(k')}$ and $\scriptN^{(k'')}$ elements, respectively, the corresponding weights $q_i^{(k)}$ satisfy

\setcounter{equation}{12}
\BEQ
q_i^{(k)}=\frac{ {\cal N}^{(k')} q_i^{(k')} +{\cal N}^{(k'')} q_i^{(k'')}} { {\cal N}^{(k')} + {\cal N}^{(k'')}}.
\label{form13}\EEQ
This is a consequence of Eq. (7) for $\scriptD_\sub^{(k')}$ and $\scriptD_\sub^{(k'')}$  with $\lambda={\cal N}^{(k')} /( {\cal N}^{(k')} + {\cal N}^{(k'')} )$
 and of the expression (10)  defining the still formal q-probabilities $q_i ^{(k)}$. 
Thus, for all possible sub-ensembles, the various final states $\scriptD_\sub^{(k)} (\tf)$ satisfy a {\it hierarchic structure} characterised by their form (\ref{fin-sub}) 
and by the additivity (13) of the q-probabilities $q_i^{(k)}$. Such an addition rule is obvious for ordinary probability distributions, and we may suspect that we are 
beginning to land in {\it standard probability theory}, but the results proved above, though suggestive, are only formal and still call for physical interpretation.

\section{Emergence of classical features}
\label{sec6}

The expressions (2) and (3) derived above are {\it the most detailed results} about ideal measurements provided by a {\it strictly formal} quantum statistical framework 
free from any interpretation, where one does not deal with individual systems but only with statistical ensembles -- possibly Gedanken, but physically consistent (Sec. 2). 
We have not only shown that the initial state $\scriptD(0)$ of S+A for a run randomly extracted from the ensemble $\scriptE$ relaxes to $\scriptD(\tf)$, but also that the 
states associated with {\it all its possible sub-ensembles} $\scriptE_\sub^{(k)}$ reach at the final time $\tf$ the equilibrium structure
$\scriptD_\sub^{(k)} (\tf)= \sum_i q_i^{(k)}\scriptD_i$ involving the {\it same building blocks} $\scriptD_i$. However, nothing yet ensures that each operator $\scriptD_i$ 
can be interpreted as a {\it final state} (1) assigned to some sub-ensemble $\scriptE_i$ yet to uncover, characterised by the outcome $A_i$.

	What remains thus to be done is to interpret the q-probabilities $q_i^{(k)} = \tr\,\scriptD_\sub^{(k)} (\tf) \hat\Pi_i$, still mathematical coefficients, as ordinary probabilities. 
In the frequency approach$^{\ref{meanings}}$, ordinary probabilities appear as numbers associated with a large ensemble and with its sub-ensembles, which have the 
following properties: they are non-negative and normalized; they are additive for disjoint sub-ensembles; they may take any value ranging from 0 to 1. Here, although 
density operators differ from probabilities because they do not refer to any sample space, the set $q_i^{(k)}$ of q-probabilities satisfy the above properties of classical 
probabilities including the hierarchic additive structure (13), except for the last one, to take any value between 0 and 1. In fact, they came out in Eq. (10) as formal objects; 
nothing ensured that their {\it range extends down to} 0 {\it and up to} 1, although nothing in the quantum formalism prevents this. In order to relate these mathematical 
objects to physical events, we ought to supplement the quantum rules of Sec. \ref{sec2} by postulating a last interpretative principle.

\subsection{Interpretation of some macroscopic q-probabilities}\label{subsec6.1}
Instead of identifying {\it any} q-probability with an ordinary probability, which would lead to paradoxes (Subsec. 2.1), we wish to introduce a much weaker principle, 
by imposing stringent conditions on the objects that will get an interpretation. We rely on the following heuristic argument. The essential feature
that distinguishes quantum mechanics from classical statistical mechanics is the
{\it non-commutative} nature of the algebra of observables. The set of projectors  $\hat\Pi_i$ 
associated with the macroscopic values $A_i$ of the pointer present in this respect a remarkable feature. Consider their commutators $[\hat \Pi_i, \hat O]$ with arbitrary 
observables $\hat O$ ($\hat O$ being bounded  when the typical dimension $G$ of the projectors becomes large). Eqs. (12a-b) imply that, in $\scriptD(\tf)$ and in 
any state $\scriptD_\sub^{(k)} (\tf)$ describing the outcome of a sub-ensemble, all q-expectation values involving commutators $[\hat\Pi_i,\hat O]$ 
have become negligible as $1/G$. 
The observables $\hat\Pi_i$ (as well as their linear combinations which describe properties of the pointer variable)  thus behave \textit{at the final time} $\tf$  {\it as if they 
commuted with the full algebra}. The quantum nature of these \textit{macroscopic variables} has become concealed as a result of the dynamics, so that they take in a
 {\it commutative behaviour at the final stage} of the process. This restrictive, quasi classical property makes the following principle natural.

\vspace{3mm}

{\it Interpretative principle 5}. Consider a set of macroscopic orthogonal projectors $\Pi_i$, a state $\scriptD$ associated at a given time with an ensemble $\scriptE$ 
 and the states $\scriptD_\sub^{(k)}$ associated with its sub-ensembles $\scriptE_\sub^{(k)}$. If the projectors have in these states the commutative behaviour expressed 
 by Eqs. (12a-b), their q-expectation values $q_i^{(k)}$ can be interpreted as physical probabilities for exclusive events, i. e., as {\it relative frequencies}\footnote{Introduced 
 here for the pointer variable to solve the measurement problem, this interpretation of q-probabilities for macroscopic quantities can be used in other contexts, such as the 
 quantum dynamics of {\it phase transitions with spontaneously broken invariance}. There, $\scriptD(t)$ denotes the state of a statistical ensemble $\scriptE$ of systems, 
 identically prepared at the macroscopic scale, and $\scriptD_i$ the equilibrium states characterised by discrete values of the macroscopic order parameter $A_i$. 
 We assume that the initial state $\scriptD(0)$ and the Hamiltonian $\hat H$ are sufficiently symmetric so as to avoid favouring the occurrence at the final time of a single 
 outcome $A_i$. We thus expect $\scriptD(t)$ to relax towards a state $\scriptD (\tf )$ of the form (2). Provided time scales are suitable, we also expect, as in 
 Sec. \ref{sec5}, the states $\scriptD^{(k)}_\sub(t)$ associated with all sub-ensembles to relax to the hierarchical structure (3). The present principle can then be used 
 to explain within quantum mechanics why, in the considered circumstances, the order parameter takes in each single experiment a well-defined value, 
 but not always the same. Implicitly assuming the principle 5, the community has rightfully not been bothered about this subtlety.\label{SolidState}}.
  
\vspace{3mm}

In the abstract formulation of quantum mechanics, arbitrary q-probabilities have no reason to be interpreted as relative frequencies (Subsec. 2.1). 
 This identification, for the specific ones $q_i^{(k)}$ submitted to the above conditions on the projectors $\Pi_i$ and on the states 
 $\scriptD_\sub^{(k)} (\tf)$, is imposed by macroscopic experience, while remaining in harmony with the quantum rules.
 
\subsection{Recovering the desired properties of ideal measurements}\label{subsec6.2} 

The above principle implies that the mathematical structure (10) of the density operators pertaining to the whole set of sub-ensembles reflects the physical structure 
of these sub-ensembles. More precisely, since the weights $q^{(k)}_i$ are interpreted as {\it standard probabilities} in the sense of frequencies, 
they can now take values ranging from 0 to 1, with 0 and 1 included.
By taking $q_j=\delta_{ij}$ in Eq. (3), we thus theoretically acknowledge the (experimentally obvious) existence of sub-ensembles $\scriptE_i$ characterised 
 by the value $A_i$ of the pointer, and to which the final state $\scriptD_i = \hat r_i \otimes \scriptR_i$ of S+A is assigned. An arbitrary sub-ensemble $\scriptE_\sub^{(k)}$ 
 can now be regarded as the merger of sub-ensembles $\scriptE_i$, each $q_i^{(k)}$ being understood as the proportion of individual runs tagged by $A_i$ in 
 $\scriptE_\sub^{(k)}$. The q-probability $p_i=\tr\,\scriptD (\tf) \hat\Pi_i$ is interpreted as the proportion in $\scriptE$ of runs having ended up with the indication $A_i$. 
 We thus recover all expected well known properties of ideal measurements.

Whether the macroscopic pointer is observed or not, the formal quantum dynamics and the above interpretative principles ensure the existence of the sub-ensembles 
$\scriptE_i$, but the latter can be explicitly identified only by reading or registering the pointer indication so as to tag the runs. Two steps are thus necessary to go from 
the initial state $\scriptD(0)$ to $\scriptD_i$. First, the irreversible dynamics of the coupled system S+A leads to $\scriptD(\tf)=\sum_i p_i \scriptD_i$ for the ensemble 
$\scriptE$, and to $\scriptD_\sub^{(k)}= \sum_i q_i^{(k)} \scriptD_i$ with unknown coefficients $q_i^{(k)}$ for its sub-ensembles. The second step, leading then to 
one of the components $\scriptD_i$, is {\it not} a consequence of some {\it evolution}, but the result of {\it selecting} the particular outcome $A_i$. It merely amounts 
 to an {\it updating of q-information} by switching from the full ensemble $\scriptE$ to the sub-ensemble $\scriptE_i$ (as in the dice example of Subsec. \ref{subsec2.4}). 

The {\it complete correlation} established by the process between the pointer indications $A_i$ and the final states $\scriptD_i$ gives access to some features 
of microscopic reality. After selection of the outcome $A_i$ and separation of the system S from the apparatus, this system is set into the quantum state 
 $\hat r_i =\tr_{\rm A} \scriptD_i =\hat\pi_i \hat r(0) \hat\pi_i / p_i$, for which the tested observable $\hat s$ has the well-defined value $s_i$ (if the eigenvalue $s_i$ is 
 non degenerate, $\hat r_i=\hat\pi_i$). We thereby derive {\it von Neumann's reduction}, which expresses the final marginal state of S for the sub-ensemble $\scriptE_i$ 
 in terms of the state $\hat r(0)$ initially assigned to S. An ideal measurement with selection of the outcome $A_i$ constitutes a {\it preparation} of the state $\hat r_i$, from which 
 we can predict the q-expectation values of all observables of S. {\it Repeating the measurement of} $\hat s$ then leaves S unchanged.

Consideration of sub-ensembles sheds light on {\it locality issues}.
Take for instance as system S a pair of particles 1 and 2 lying far apart and carrying
 the spins $\hat {\bf s}^{(1)}$ and $\hat {\bf s}^{(2)}$, initially prepared in the singlet state 
 $\hat r(0) = \half (|\!\up\down\rangle \\ - |\!\down\up\,\rangle\,)(\,\langle \up\down\! |-\langle \down\up\! | \,)$.
  The measured observable is the $z$-component $\hat s^{(1)}_z$ of the spin 1. The pointer has two possible outcomes $A_\up$ and $A_\down$ 
 associated with $s^{(1)}_z =1$ and $s^{(1)}_z=-1$, and their selection at the time $\tf$ produces two sub-ensembles $\scriptE_\up$ and $\scriptE_\down$
 of runs for which the reduced states of S are $\hat r_\up=|\hspace{-0.6mm}\up\down\rangle\,\langle \up\down\hspace{-0.6mm}\!|$ 
and  $\hat r_\down=|\hspace{-0.6mm}\down\up\rangle\,\langle\down\up\hspace{-0.6mm}\!|$, respectively. 
 The interaction $\hat H_{\rm SA}$ is localized in the vicinity of the particle 1 and is switched on during the time lapse $0 < t < t_{\rm decoup}$
  (with $t_{\rm decoup} < \tf$). The fact that the particle 2 lies beyond the range of the apparatus is consistent with the time-invariance, 
  for the full ensemble $\scriptE$, of its marginal state 
  $\hat r^{(2)}(t)=\half(\, |\!\up\rangle\,\langle \up\!| + |\!\down\rangle\,\langle \down\!| \,)$.
  However, for the sub-ensemble $\scriptE_\up$ of runs, one can assign to S the reduced state 
  $\hat r_\up=|\hspace{-0.6mm}\up\down\rangle\,\langle \up\down\hspace{-0.6mm}\!|$ 
  after the reading time $\tf$, but also already after the decoupling time $t_{\rm decoup}$, since S cannot evolve after $t_{\rm decoup}$. 
  Likewise, one can assign to the spin 2 the reduced marginal state 
  $\hat r^{(2)}_\up=|\hspace{-0.6mm}\down\rangle\,\langle \down\hspace{-0.6mm}\!|$ 
at any time $t > 0$. The change of state of the spin 2 from $\hat r^{(2)}$  (for $\scriptE$) to  $\hat r^{(2)}_\up$ (for  $\scriptE_\up$ ), 
which takes place far from the measuring apparatus, is evidently not a result of some non-local physical effect; it is merely a {\it non-local inference}
by the experimenter, based on his knowledge of the initial intricate state $\hat r(0)$ and his possibility 
to select the runs of the sub-ensemble $\scriptE_\up$  by a retroactive use of information gathered through the pointer. 
If experiments involving the spin 2 are performed at an arbitrary time $t > 0$ (to determine for instance a q-correlation between 1 and 2),
 they can be analyzed by assigning to this spin the state $\hat r^{(2)}_\up=|\hspace{-0.6mm}\down\rangle\,\langle \down\hspace{-0.6mm}\!|$
 for the runs belonging to  $\scriptE_\up$  and the state $\hat r^{(2)}_\down=|\hspace{-0.6mm}\up\rangle\,\langle \up\hspace{-0.6mm}\!|$
 for the runs belonging to  $\scriptE_\down$.  However, the very sorting of runs requires observation of the pointer
  and transfer of this information towards the processing point;  the preparation of the spin 2 in either 
 the state $\hat r^{(2)}_\up$  or the state $\hat r^{(2)}_\down$  through measurement of $\hat s^{(1)}_z$ can therefore be acknowledged only after the time $\tf$.
Altogether, nonlocality lies only in the q-correlations between the two spins 1 and 2 that exist in their initial state; 
these two parts do not communicate later on. All physical processes involved in the measurement are local.

The uniqueness of the outcome of {\it individual runs} for an ideal measurement process also emerges theoretically from the identification of the weights $q_i^{(k)}$ 
 as frequencies, since we can characterise after the process a single compound system S+A belonging to $\scriptE_i$ by the state $\scriptD_i$. 
 A  {\it dynamical  solution of the measurement problem} (Subsec. 1.1) has thus come out.
 
In general, qualitatively new physical properties emerge in a change of scale, and their theoretical explanation goes through some interpretative principle 
which complements the formalism. For instance, in statistical mechanics, the principle 2 of Sec. 2.3 is used to explain how macroscopic continuity of matter 
emerges from a discrete microscopic structure, or how irreversibility emerges from reversible equations of motion. Here, the qualitative changes that result 
from the macroscopic size of the apparatus concern not only {\it phenomena} (the measurement process is irreversible), but also, remarkably, {\it concepts}: 
Classical features emerge from a merely formal quantum approach supplemented by the interpretative principle 5 which concerns {\it only the pointer} 
of the apparatus.

\subsection{Restricted field of the interpretative principle 5}\label{subsec6.3}

The principle 5 of Subsection 6.1 cannot be extended carelessly, as it is founded on several stringent requirements. 

($i$) The {\it effective commutation} of the projectors $\hat\Pi_i$ with the full algebra relies on the {\it macroscopic character} of these projectors, since Eq. (12b) 
 holds only for a macroscopic pointer ($G\gg 1$). More generally, we expect the same ideas to hold for macroscopic systems involving several equilibrium states 
 distinguished through high-dimensional projectors, for instance systems with broken discrete invariance$^{\ref{SolidState}}$ ($A_i$ is then replaced by the value 
 of the order parameter). 

($ii$) Moreover, this effective commutativity of the projectors $\hat\Pi_i$ is ensured {\it only at the final time}, as Eqs. (12a-b) involve the final states $\scriptD_\sub^{(k)}$. 
 During the process, the non-Abelian nature of $\hat\Pi_i$ cannot be neglected since the pointer must evolve from its initial metastable state to one of the stable states, 
 and this time-dependence of $\tr\,\scriptD(t) \hat \Pi_i$ requires that $[\hat\Pi_i, \hat H]$  is effective until equilibrium is reached.
 Note that, whereas the projector $\hat\Pi_i$ pertaining to A 
 does not commute with $\hat H$ ($\hat\Pi_i$ is effectively conserved only after $\tpf$), the projector $\hat\pi_i$ pertaining to S commutes with $\hat H$,  
 so that ${\rm tr}\, \scriptD(t) \hat \pi_i$ remains constant at all times. However, being microscopic, $\pi_i$ cannot satisfy relations such as (12b), 
 and the principle 5 does not apply to it.

($iii$) The consideration of {\it sub-ensembles} has also been essential. If we wish individual runs to provide the outcomes $\scriptD_i$, the necessary conditions (3) 
 must be fulfilled. Due to the existence of incompatible decompositions of $\scriptD(\tf)$ (Subsec. \ref{subsec5.1}), it is not justified to postulate {\it directly}, as generally 
 done, that the coefficients $p_i$ in $\scriptD(\tf)=\sum_i p_i\scriptD_i$ might be interpreted as frequencies of the outcomes $\scriptD_i$ in the full ensemble $\scriptE$:
 this fallacy is the measurement problem that we addressed. We escaped this loophole (Sec. 5.4), by eliminating the quantum ambiguity through a dynamical process, 
 the poly-microcanonical relaxation  \cite{allahverdyan2013understanding}, which provides the expected structure for the states $\scriptD_\sub^{(k)}$.

\subsection{Status of Born's rule}\label{subsec6.4}

The relative frequency $p_i$ of occurrence of the macroscopic value $A_i$ {\it of the pointer} has been found, according to the above principle, as 
 $p_i={\rm tr}\, \scriptD(\tf) \hat\Pi_i$ in terms of the {\it final state} of S+A. However, such a proportion is currently expressed by Born's rule 
 $\tr_{\rm S} \hat r(0) \hat\pi_i$, which disregards the apparatus and involves only the {\it initial state} of the {\it tested system}. In the light 
 of the restrictions about principle 5, we are not entitled in the present approach to directly interpret the q-probability $\tr_{\rm S} \hat r(0) \hat\pi_i$ 
 as a genuine probability and to admit blindly Born's rule. To derive it theoretically, we need to rely on the following two properties. 

($i$) In the final state (2) of S+A, the marginal state $\hat r_i$ of S is {\it fully correlated} with the macroscopic indication $A_i$. Using the identity 
 ${\rm tr}_{\rm A} \scriptR_i \hat \Pi_j= \tr_{\rm S} \hat r_i \hat \pi_j = \delta_{ij }$, we can thus identify  ${\rm tr}\, \scriptD(\tf) \hat\pi_i$ with the true 
 probability $ p_i={\rm tr}\, \scriptD(\tf) \hat\Pi_i$. However, this feature was not granted a priori, as it results from the dynamics of the process. 
 Indeed, if the coupling $\hat H_{\rm SA}$ is {\it too weak}, the ``registration'' process considered in Sec. \ref{sec4} may be imperfect, driving $\scriptR_{ii}(t)$ 
  with some probability to a wrong equilibrium state $\scriptR_j$ with $j\neq i$ (see  ref. \cite{allahverdyan2013understanding}, sec. 8).  
  The resulting imperfection of the correlation between $s_i$ 
  and $A_i$ then produces a {\it violation of Born's rule}, with a q-probability ${\rm tr}\, \scriptD(\tf) \hat\pi_i$ of $s_i$ in the final state different from the observed 
  frequency tr $\scriptD(\tf) \hat\Pi_i$ of $A_i$.
 
($ii$) The {\it conservation law} $[\hat H, \hat s]=0$ implies the relation 
  ${\rm tr}\, \scriptD(\tf) \hat\pi_i = {\rm tr}\, \scriptD(0) \hat\pi_i = {\rm tr}_{\rm S} \hat r(0) \hat\pi_i$ 
  between q-probabilities of $s_i$  at the initial and final times, again as a {\it consequence of the dynamics of S+A} during the measurement process. 
 With the above property $p_i\equiv \tr \scriptD(\tf) \hat\Pi_i =\tr \scriptD(\tf) \hat\pi_i$,
 this finally leads to Born's expression $p_i=\tr_{\rm S} \hat r(0) \hat\pi_i$.

The complete correlation between $s_i$ and $A_i$ allows us to extend the ordinary probabilistic interpretation of $p_i={\rm tr}\, \scriptD(\tf) \hat\Pi_i$, 
issued from the principle 5, to some microscopic quantities. In the present approach,
we may for instance determine from $p_i$ the ordinary expectation value of $\hat s$ and its variance, 
or write the standard conditional probability of $\hat s$ to be equal to $s_j$ if the pointer takes the value $A_i$ as  
$\tr\,\scriptD(\tf) \hat\pi_j \hat\Pi_i / \tr\,\scriptD(\tf) \hat\Pi_i =\delta_{ij}$.
In fact, we have shown that such identifications are licit {\it only at the final time}, 
when ordinary probabilities have emerged {\it after interaction with an apparatus} designed to measure the observable $\hat s$. 
The occurrence of the {\it initial state} in Born's expression $p_i= \tr_{\rm S} \hat r(0) \hat\pi_i$ is somewhat misleading. 
Although one can infer from the measurement of $\hat s$ some {\it formal} properties of $\hat r(0)$, 
one should not interpret it as a {\it probability} of $\hat s$ to take the value $s_i$ in the initial state $\hat r(0)$ of S (Subsec. 7.3).
We have avoided such an over-interpretation of the quantum formalism, which would lead to the logical contradictions exemplified by 
Bell's inequalities or the GHZ paradox.

\section{Epilogue: Quantum mechanics as a half-blind theory }
\label{sec7}

The above reasoning appears as the complete opposite of a recent approach \cite{auffeves2014contexts,auffeves2015simple} 
which introduces as a starting point some physical axioms pertaining to the system S placed in all imaginable contexts. 
There, an ordinary probabilistic description applies for each context. Gleason's theorem is then used to unify the contexts
\cite{gleason1957measures, busch2003quantum,caves2004gleason}, and thus to construct for S the standard
mathematical formalism of quantum mechanics. Here, we conversely start from this abstract formalism. 
We consider for S a single context which is materialised by a macroscopic apparatus A, and analyse the dynamics of the compound quantum system S+A. 
The properties of the measurement emerge at the end of this process owing to the introduction of a few physical principles. 
We comment below the main features of the present approach.

\subsection{A minimalist and macroscopic interpretation}
\label{subsec7.1}

Measurement theory is often treated in close connection with interpretation of quantum mechanics. Our scope here has been more limited. We did not attempt to 
interpret the quantum formalism taken as granted, but only proposed an interpretation of ideal quantum measurement processes. We followed two paths in parallel. 
On the formal side, 
we discussed which features of the Hamiltonian are needed to ensure that the process has all required properties, and we brought out the most detailed results that 
quantum dynamics (without interpretation) can provide. On the conceptual side, we looked for the least numerous and narrowest possible interpretative principles 
needed to establish, for ideal measurements, a bridge between formal quantum results and physical reality. All other mathematical objects manipulated in quantum 
theory, the operator-valued observables and states as well as their scalar products, the ``q-expectation values'', have remained abstract. 

Thus, only the final indications of the macroscopic pointer were eventually described by means of ordinary probabilities for individual runs. Some microscopic physical 
properties selected by the process could subsequently be grasped as the result of an inference. Indeed, most interpretative principles introduced above in a natural way 
refer to macroscopic properties. They lie {\it astride macrophysics and microphysics} and are consistent both with our macroscopic experience and with the quantum formalism.       

Some of these principles are minimalist, in the sense that they are submitted to drastic conditions, which however are sufficient for our purpose. The principles 1 and 5, 
which identify some q-probabilities with relative frequencies, do not apply to arbitrary observables and states, but only to particular macroscopic observables and particular 
states satisfying stringent conditions (Subsec. 6.3). The principle 4 helped us to explain through a sub-ensemble analysis the apparent ``bifurcation'' (or ``multifurcation'') 
which leads from the single initial state $\scriptD(0)$ to several final states $\scriptD_i$. However, this principle introduced such sub-ensembles only by the end of the 
measurement process, after the time $\tpf$ at which $\scriptD(t)$ had already reached the form $\sum_i p_i \scriptD_i$. In fact, the possibility of recognising physical  
sub-ensembles within $\scriptD$ exists only at the last stage of the measurement process and \textit{emerges from the dynamics} of A.

\subsection{Measurement as transfer of q-information and promotion to real information}
\label{subsec7.2}

The principles 2 (Subsec. 2.3) and 3 (Subsec. 3.2) are consistent with the conception of a quantum state as a {\it catalogue of q-information} referring to a statistical ensemble. 
In this prospect, dynamics produce transfer of q-information within the set of observables; selection of a sub-ensemble affords updating of q-information. An ideal 
measurement appears as a {\it processing of information}, which involves transformations of the q-information carried formally by quantum states, and conversion of 
q-information into ordinary information accessible to experiment\footnote{States being viewed as catalogues of knowledge (Sec. 2), q-information about S is updated 
in an ideal measurement by replacing the initial state $\hat r(0)$ by $\hat r_i$ if $A_i$ is selected, or by $\sum_i p_i \hat r_i$ if the indications  of A are not selected. 
If the tested observable is not fully specified, the least biased subsequent predictions should rely on a state obtained by averaging over all possible interaction processes. 
If for instance, one is aware that an ensemble of spins initially prepared in the state $\hat r(0)$ have been measured in some direction, but if one knows neither in which 
direction nor the results obtained, one should assign to the final state the density operator $\frac{1}{3}[\hat 1+\hat r(0)]$ as being the best (but imperfect) 
description. (To show this, write $\hat r(0)$ in its polar form and then as the projected form after a measurement.)\label{old12}}. Such changes can be made quantitative by evaluation of the relevant von Neumann entropies. 
Let us review these informational aspects of the above treatment. 

The initial metastable density operator $\scriptR(0)$ of A, defined by some macroscopic data, is provided by the principle 3 as the least informative one that accounts 
for these data (Section 3), while $\hat r(0)$ encodes the q-information characterising initially the ensemble $\scriptE$ in which S is embedded. The relaxation of the coupled 
system S+A (Section 4) consists in a transfer of q-information from some degrees of freedom to others, with possible loss towards inaccessible ones (principle 2). 
Thus, in truncation, q-information leaks towards inaccessible q-correlations between S and an increasingly large number of degrees of freedom of A (with a huge 
recurrence time). In registration, the loss of q-information towards the bath is partly compensated for by the creation of complete q-correlations between $s_i$ and $A_i$. 
Some downgrading is also produced for each sub-ensemble $\scriptE_\sub^{(k)}$ by the poly-microcanonical mechanism of relaxation (Sec. 5).   

The principle 5 (Sec. 6.1) finally expresses that the q-information about the pointer variable, embedded within the state $\scriptD(\tf)$, can be converted into 
{\it ordinary, readable information}, and disclosed after the final time in the form of relative frequencies $p_i$ of occurrence of $A_i$. As usual, selection of the 
sub-ensembles $\scriptE_i$, tagged by the value $A_i$ of the pointer, increases q-information.  

As regards the system S itself, some ordinary information pertaining to the tested observable $\hat s$ has thus been extracted owing to the correlations between 
$s_i$ and $A_i$ built up dynamically by the coupling with the apparatus. The dynamical process undergone by S+A has pulled out from the
{\it latent q-information  contained in} $\hat r(0)$ 
the part associated with $\hat s$, and has {\it converted it into true information}. The initial q-informations about observables that commute with $\hat s$ have been 
preserved; they are encoded within the reduced states $\hat r_i$, which may subsequently be used for further experiments. 

However, gaining information on S through a quantum measurement \textit{requires an irreversibility} of the physical process of interaction between S and A, 
hence a loss of q-information. Not only does this loss take place within A, but von Neumann's reduction expresses the vanishing after ideal measurement of the 
q-expectation values $\tr_{\rm S} \hat r_i \hat O_\off$ of {\it all off-diagonal observables} $\hat O_\off$ of S such that $\hat\pi_i \hat O_\off \hat\pi_i = 0$ 
for all $i$. Remarkably, 
gaining full information about $\hat s$ requires perturbing S so as to destroy the whole q-information about the observables that do not commute with $\hat s$. 
This unavoidable {\it loss of q-information} about the observables of S incompatible with $\hat s$ is a {\it price to pay for testing the quantum observable} $\hat s$.

\subsection{Role of the apparatus}\label{subsec7.3}

The apparatus plays a major role, not only experimentally but also in the theory of ideal measurements. As usual in statistical mechanics, it is owing to the macroscopic 
size of the apparatus that the irreversibility of the measurement process emerges from the reversible microscopic dynamics (Secs. 4 and 5). It is also this macroscopic 
size which produces at our scale other remarkable types of emergence of features qualitatively different from those of quantum theory (Sec. 6).

Technically, we have seen that the dynamical equations (6) which govern the relaxation of S+A for the full ensemble of runs are expressed only in terms of the apparatus. 
The tested system only appears through the factors $\hat h_i$ of the coupling $\hat H_{\rm SA}$ which trigger the evolution of A towards one or another of its stable states. 
The system S does not even intervene at all in the poly-microcanonical relaxation that takes place for the sub-ensembles (Sec. 5).

As regards the interpretation of the measurement outcomes, most principles that we have been led to introduce also concern {\it only the macroscopic apparatus}. 
We have stressed that the probability $p_i$ refers to the pointer observable, and is associated only indirectly with the eigenvalues of $\hat s$ through the full 
correlation between S and A.

\subsection{Most q-probabilities should remain uninterpreted}\label{subsec7.4}

Due to the omnipresence of the apparatus in the analysis of ideal quantum measurements, their outcomes should {\it not be viewed as intrinsic properties} of the 
system S irrespective of A, but as {\it joint properties} of S and A. In particular, the relative frequencies of occurrence of the pointer indications $A_i$ 
came out theoretically as $p_i={\rm tr} \,\scriptD(\tf) \hat\Pi_i$. This expression has pre-eminence over Born's formula $ p_i={\rm tr}_{\rm S} \hat r(0) \hat\pi_i$, 
which although important is only a by-product of the dynamics of S+A (Subsec. 6.4) and has no fundamental character,
as is obvious when the measurement is imperfect.

We also stressed that one should not be misled by the occurrence of the initial state $\hat r(0)$ in Born's rule. Retrodiction from the outcomes of measurements 
 towards properties of S at the initial time is legitimate only for abstract q-probabilities, and $ p_i$ {\it should not be interpreted as a true probability} for $\hat s$ 
 to take the value $s_i$ in the state $\hat r(0)$, as is often taught. 

For instance, in experiments testing Bell's inequalities, spin pairs are all prepared similarly in a given initial state $\hat r(0)$ of S, and {\it several series of measurements} 
 are performed, each one using a pair of detectors oriented in two given directions. Each single run provides a value +1 or -1 for the product of the spin components in 
 the considered directions. For {\it each} such setting, one gets from the ensemble of runs a correlation between the two spins, 
 which it is legitimate to interpret as a \textit{true correlation}, but only {\it in the final state}. 
 However, by retrodiction towards the initial state $\hat r(0)$, one may interpret this quantity only as an {\it abstract q-correlation}, not as a 
 true correlation. Indeed, putting together such q-correlations issued from different series of measurements violates Bell's inequalities, which should be satisfied by 
 true physical correlations. As a quantum measurement is a joint property of S and A, we are not allowed to interpret {\it simultaneously} as real properties of the initial state 
 of S the results of experiments obtained with {\it different apparatuses} (here with different directions of the detectors).
 This deep property of quantum measurements is in line with the absence, for quantum states, of a sample space as in ordinary probability theory
  \cite{hess2005bell,khrennikov2007bell,hess2009possible,nieuwenhuizen2011contextuality,kupczynski2015bell}.

The situation is even worse with the GHZ paradox, experimentally verified.
There, complete q-correlations between several observables are exhibited by measurements, performed on identically prepared systems S 
and involving {\it different apparatuses} (hence different ensembles of joint systems S+A). If one then combines some identities implied by these 
q-correlations, using elementary algebraic rules that would hold for ordinary correlations, one stumbles on a 
{\it logical contradiction}\footnote{The 3-spin example of the GHZ paradox involves several observables $\hat a^{(j)}$, $\hat b^ {(j)}$  ($j=1,2,3$) with 
$\hat b^{(3)}\equiv  \hat b^{(1)} \hat b^{(2)}$, defined by $\hat a^{(1)} = \hat \sigma^{(1)}_x$ , $\hat b^{(1)} = \hat \sigma^{(2)}_z \hat \sigma^{(3)}_z$ , etc.
 All these operators have eigenvalues $a^{(j)}= \pm 1$, $b^{(j)}= \pm1$ and commute with each other, 
except for $\hat a^{(j)}$ and $\hat b^{(k)}$ which anticommute when $j\neq k$. 
The system is prepared in a pure state, the common eigenstate of $\hat a^{(j)} \hat b^{(j)}$ ($j=1,2,3$) with eigenvalues 1.
At the formal level, this yields the q-expectation values $\langle \hat a^{(j)} \rangle = \langle\hat b^{(j)}\rangle = 0$ and the complete q-correlations 
$\langle \hat a^{(j)} \hat b^{(j)}\rangle =1$ ($j = 1, 2, 3$), $\langle \hat a^{(1)} \hat a^{(2)} \hat a^{(3)} \rangle = - 1$. 
Physically, the property $\langle \hat a^{(1)} \hat b^{(1)}\rangle =1$ can be tested through
simultaneous measurements of the commuting observables $\hat a^{(1)}$ and $\hat b^{(1)}$;
each run provides a fully correlated outcome $a^{(1)}$, $b^{(1)}$ such that $a^{(1)} = b^{(1)}$, in agreement with $\langle \hat a^{(1)}\hat b^{(1)}\rangle=1$. 
Likewise, {\it other} sets of measurements provide outcomes satisfying $a^{(2)} = b^{(2)}$ and $a^{(3)} = b^{(3)}$, 
while simultaneous measurement of $\hat a^{(1)}$, $\hat a^{(2)}$, $\hat a^{(3)}$ yields $a^{(3)} =  - a^{(1)} a^{(2)}$ for each run. 
However, accounting for $\hat b^{(3)} = \hat b^{(1)} \hat b^{(2)}$ and naively combining the identities $a^{(1)} = b^{(1)}$, 
$a^{(2)} = b^{(2)}$ and $a^{(3)} = b^{(3)}$ would yield $a^{(3)} = + a^{(1)} a^{(2)}$, in flagrant contradiction with the quantum result 
$\langle \hat a^{(1)} \hat a^{(2)} \hat a^{(3)} \rangle = -1$, issued from non-commutation. 
Note that the incompatibility between the relations $a^{(1)} = b^{(1)}$, $a^{(2)} = b^{(2)}$, $a^{(3)} = b^{(3)} = b^{(1)} b^{(2)}$ and 
$a^{(3)} = - a^{(1)} a^{(2)}$, each of which is satisfied by a measurement performed with a specific apparatus, 
holds even if a single run is considered for each of the four measurements.}  \cite{greenberger1989going}:
one would find a result $+1$ instead of the actual quantum value $-1$. 

If we wish to understand quantum mechanics within standard probability theory and standard logic, we ought to keep microscopic q-expectation 
values uninterpreted,  except at the end of a measurement. Although q-probabilities, q-expectation values or q-correlations are 
{\it mathematically expressed in terms of} S {\it alone}, they acquire a consistent 
{\it physical meaning only in the presence of a dedicated measurement apparatus}. 
Probabilistic or logical paradoxes occur when results obtained in {\it different experimental contexts} are interpreted as ordinary 
 expectation values or correlations and are put together.

\vspace{3mm}
 
Attributing a complete interpretation, of some kind or another, to the so-called ``states'' or ``expectation values'' of quantum theory thus appears not only unnecessary
but even troublesome. An abstract formulation is advocated, especially for tutorial purposes, where interpretation is restricted to measurements outcomes. In particular, it is 
advisable to avoid misleading vocabulary and to carefully distinguish q-probabilities from ordinary probabilities. Such a distinction is currently made for q-bits and bits 
in quantum computation; there one stresses that readable bits may be produced only by partly destroying q-bits, which cannot fully be grasped. Likewise, for arbitrary 
observables $\hat O$, one should discriminate when teaching quantum mechanics q-expectation values $\tr \,\scriptD\hat O$, which are latent mathematical objects, 
from true expectation values $\tr \,\scriptD\hat s$, which emerge at the issue of the measurement of an observable $\hat s$. Distinguishing formal q-probabilities, 
which characterise quantum ``states'', from ordinary probabilities, which govern data issued from measurements, would help to understand the status of Born's rule, 
and to circumvent apparent contradictions that arise when one combines q-correlations which cannot be measured 
with a single experimental measurement setting.

\section*{Acknowledgment} 
It is a great pleasure to acknowledge extensive, passionate and thorough discussions with Franck Lalo\"e.

\section*{Appendix A. Super-systems, q-expectation values vs average values, and maximum entropy}
 
Inspired by the equivalence, in ordinary probability theory, between expectation values and average values, we wish here to compare q-expectation values with quantum 
averages over a large number of samples. To this aim, we introduce a large set of systems $\rS^{[n]}$ ($n=1, 2,\cdots, \scriptN$) similar to the (small or large) system S 
of interest \cite{balian1987equiprobability}. We regard the merger of $\rS \equiv \rS^{[1]}$ with all its siblings $\rS^{[2]}, \cdots, \rS^{[\scriptN]}$ as a large compound 
{\it super-system} \bS {}\,  =  $\{\rS^{[1]} , \rS^{[2]},\cdots , \rS^{[\scriptN]}\}$. One should not confuse the super-system \bS\,{}, which is a single compound Gedanken 
system, with the ensemble $\scriptE$ to which S belongs. (In fact, the quantum description of  \bS \  involves a ``super-ensemble'' 
$\uline{\scriptE}$ of copies of \bS\,{}.) 

With each observable $\hat O\equiv\hat O^{[1]}$ in the Hilbert space $\scriptH^{[1]}$ of $\rS\equiv \rS^{[1]}$, we associate the {\it average observable} \bOhat \, = 
$\scriptN^{-1}\sum_n \hat O^{[n]}$  in the Hilbert space $\Pi_n\otimes\scriptH^{[n]}$ of \bS.  (Each term of this sum is meant as the tensor product of $\hat O^{[n]}$ 
by all the unit operators associated with the other systems $\rS^{[n']}$ with $n'\neq n$.) The various average observables, pertaining to the macroscopic super-system 
\bS, nearly commute with one another, as the commutation relation $[\hat O_1,\hat O_2]=i\hat O_3$ for S implies
 [\bOhat$\,_1,$ \bOhat$\,_2]=i$\bOhat$\,_3/\scriptN$ for \bS\,, with $\scriptN\gg1$. (The set of average observables constitute a Lie algebra, but not a full algebra 
 since their products lie outside their set.) 
 Accordingly, these average observables behave quasi-classically in the large $\scriptN$ limit and may be assigned simultaneously rather well defined values.

Let us consider a state \bDhat \,\,  of \bS,  invariant under permutations of the subsystems $\rS^{[n]}$. These subsystems $\rS^{[n]}$ all lie in the same marginal 
state $\scriptD$, and the q-expectation values tr $\scriptD\hat {O}$ for S and \uline{Tr} \bDhat\,\,\bOhat {}\,  \, for \bS \, are equal. If the correlations between 
subsystems are weak, of order less than $\scriptN^{-1}$,  while the q-variance of $\hat O$ in the state $\scriptD$ is finite,  the q-variance of \bOhat \, for the 
super-system \bS, i.e.,  \uline{Tr} \bDhat\,\,\bOhat\ $^2 -$ (\uline{Tr} \bDhat\,\,\bOhat\,)$^2 \sim [$tr $\scriptD\hat O^2-$ (tr $\scriptD\hat O)^2$]$/\scriptN$, is 
negligible in relative size. We can thus apply the principle 1 of Subsec. \ref{subsec2.2} to the average observable \bOhat \,  
of the large supersystem \bS, and interpret the q-expectation value \underline{Tr} \bDhat\,\,\bOhat {}\, as an ordinary value. This leads us to identify 
a {\it formal q-expectation value} $\langle\hat O\rangle$ for the (possibly small) system S with the macroscopic value of the corresponding 
{\it average observable} \bOhat \, \, for the super-system \bS. 

Though somewhat artificial due to the virtual nature of the supersystem \bS, the latter identification was a key point in a general proof \cite{balian1987equiprobability} 
 of the {\it maximum von Neumann entropy criterion} based upon {\it Laplace's indifference or equiprobability principle}. The derivation extended a classical 
 argument by Gibbs, who had shown that, for given  $\langle \hat H\rangle$, a microcanonical equilibrium state for the super-system \bS \ entails for the 
 (possibly small) system S a canonical state. A similar purpose in quantum mechanics (Subsec. \ref{subsec3.2}) is to assign the least biased state to a quantum 
 system S when the sole q-expectation values $\langle  \hat O_p  \rangle$ of some observables  $\hat O_p$  of S are given ($1\le p\le p_{\rm max}$). 
 The quantities $\langle  \hat O_p  \rangle$ are identified with the values $\langle$\bOhat$\,_p \rangle$ of the average observables \bOhat$\,_p$  of the 
 associated super-system \bS, which nearly commute and present small fluctuations. As the macroscopic data $\langle $\bOhat$\,_p \rangle$ for the 
 supersystem \bS \ are defined within a small margin (like the energy for the microcanonical state), many kets are compatible with them for large $\scriptN$, 
 and unitary invariance in the Hilbert space of \bS \ sets these kets on the same footing. Laplace's indifference principle then leads to assign to \bS \ 
 a density operator concentrated over these kets, which generalises the microcanonical state. The corresponding density operator $\scriptD$ 
 of S then results by tracing out from \bS \ its subsystems S$^{[n]}$ with $2\le n \le \scriptN$. Such a program, which presents difficulties when the observables $\hat O_p$ 
 do not commute, has been achieved in Ref. \cite{balian1987equiprobability}\footnote{The proof requires that the width of the generalized microcanonical distribution of the 
 super-system behaves as $N^{-\alpha}$ with $1/2 < \alpha \le 3/4$. The lower bound on this width allows one to by-pass the argument 
  raised against the macroscopic approach to quantum probabilities in the discussions of Refs.  \cite{squires1990alleged,aharonov2002macroscopic,finkelstein2003comment} .
  This criticism was based on ``microscopic definiteness'' (see footnote 8), while here the system cannot be not close to an eigenstate.\label{fnAppA}}. It provides for $\scriptD$
  the exponential of a weighted sum of the observables $\hat O_p$, the same result as the outcome of the maximisation 
 of von Neumann's entropy under constraints on $\langle  \hat O_p  \rangle$. The principle 3 of subsection \ref{subsec3.2} may therefore be replaced by 
 Laplace's indifference principle, used in connection with the equivalence between q-expectation values and ensemble averages 
 (principle 1 of Subsec. \ref{subsec2.2}), and with unitary invariance.

\vspace{3mm}



\begin{thebibliography}{10}
\expandafter\ifx\csname url\endcsname\relax
  \def\url#1{\texttt{#1}}\fi
\expandafter\ifx\csname urlprefix\endcsname\relax\def\urlprefix{URL }\fi
\expandafter\ifx\csname href\endcsname\relax
  \def\href#1#2{#2} \def\path#1{#1}\fi

\bibitem{heisenberg1925uber}
W.~Heisenberg, \"Uber quantentheoretische Umdeutung kinematischer und
  mechanischer Beziehungen, Zeitschrift f\"ur Physik 33 (1925) 879--893.

\bibitem{wheeler2014quantum}
J.~A. Wheeler, W.~H. Zurek, Quantum theory and measurement, Princeton
  University Press, 2014.

\bibitem{alter2001quantum}
O.~Alter, Y.~Yamamoto, Quantum measurement of a single system, Wiley, New York,
  2001.

\bibitem{braginsky1995quantum}
V.~B. Braginsky and F.~Y. Khalili, 
Quantum measurement, Cambridge University Press, 1995.

\bibitem{home1992ensemble}
D.~Home, M.~A. Whitaker, Ensemble interpretations of quantum mechanics. a
  modern perspective, Physics Reports 210~(4) (1992) 223--317.

\bibitem{namiki1993quantum}
M.~Namiki, S.~Pascazio, Quantum theory of measurement based on the
  many-Hilbert-space approach, Physics Reports 232~(6) (1993) 301--411.

\bibitem{zurek2003decoherence}
W.~H. Zurek, Decoherence, einselection, and the quantum origins of the
  classical, Reviews of Modern Physics 75~(3) (2003) 715.

\bibitem{sewell2007can}
G.~Sewell, Can the quantum measurement problem be resolved within the
framework of Schr\"odinger dynamics?, Mark. Proc. Rel. Fields 13 (2007) 425.

\bibitem{laloe2012we}
F.~Lalo{\"e}, Do we really understand quantum mechanics?, Cambridge University
  Press, 2012.

\bibitem{allahverdyan2013understanding}
A.~E. Allahverdyan, R.~Balian, T.~M. Nieuwenhuizen, Understanding quantum
  measurement from the solution of dynamical models, Physics Reports 525~(1)
  (2013) 1--166.

\bibitem{narnhofer2014reduction}
H.~Narnhofer, W.~F. Wreszinski, On reduction of the wave-packet, decoherence,
  irreversibility and the second law of thermodynamics, Physics Reports 541~(4)
  (2014) 249--278.

\bibitem{weinberg2016happens}
S.~Weinberg, What happens in a measurement?
 Physical Review A 93~(3) (2016)  032124.

\bibitem{demuynck2006foundations}
W.~M. de~Muynck, Foundations of quantum mechanics, an empiricist approach, Vol.
  127, Springer Science \& Business Media, 2006.

\bibitem{demuynck2016crucial}
W. M. de Muynck, On the crucial role of measurement in the interpretation of quantum mechanics (preprint available on ResearchGate, 2016).  

\bibitem{luders1950zustandsanderung}
G.~L{\"u}ders, {\"U}ber die Zustands{\"a}nderung durch den Me{\ss}proze{\ss},
  Annalen der Physik 443~(5-8) (1950) 322--328.

\bibitem{luders2006concerning}
K.~A. Kirkpatrick, Concerning the state-change due to the measurement process,
  Annalen der Physik 15~(9) (2006) 663--670.

\bibitem{home2013conceptual}
D.~Home, Conceptual foundations of quantum physics: an overview from modern
  perspectives, Springer Science \& Business Media, 2013.

\bibitem{schrodinger1935discussion}
E.~Schr{\"o}dinger, Discussion of probability relations between separated
  systems, in: Mathematical Proceedings of the Cambridge Philosophical Society,
  Vol.~31, Cambridge Univ. Press, 1935, pp. 555--563.

\bibitem{schrodinger1936probability}
E.~Schr{\"o}dinger, Probability relations between separated systems, in:
  Mathematical Proceedings of the Cambridge Philosophical Society, Vol.~32,
  Cambridge Univ. Press, 1936, pp. 446--452.

\bibitem{park1988thermodynamic}
J.~L. Park, Thermodynamic aspects of Schr{\"o}dinger's probability relations,
  Foundations of physics 18~(2) (1988) 225--244.

\bibitem{allahverdyan2003curie}
A.~E. Allahverdyan, R.~Balian, T.~M. Nieuwenhuizen, Curie-Weiss model of the
  quantum measurement process, Europhysics Letters 61~(4) (2003) 452.

\bibitem{thirring2013course}
W.~Thirring, A course in mathematical physics 3: Quantum mechanics of atoms and
  molecules, Springer Science \& Business Media, 2013.

\bibitem{david2015quantum}
F.~David, 
The formalisms of quantum mechanics, an introduction. Lecture notes in physics 893, Springer, 2015.

\bibitem{vankampen1988ten}
N.~Van~Kampen, Ten theorems about quantum mechanical measurements, Physica A:
  Statistical Mechanics and its Applications 153~(1) (1988) 97--113.

\bibitem{balian1999incomplete}
R.~Balian, Incomplete descriptions and relevant entropies, American Journal of
  Physics 67~(12) (1999) 1078--1090.

\bibitem{squires1990alleged}
E.~J. Squires, On an alleged 'proof' of the quantum probability law, Physics
  Letters A 145~(2) (1990) 67--68.

\bibitem{aharonov2002macroscopic}
Y.~Aharonov, B.~Reznik, How macroscopic properties dictate microscopic
  probabilities, Physical Review A 65~(5) (2002) 052116.

\bibitem{finkelstein2003comment}
J.~Finkelstein, Comment on ``How macroscopic properties dictate microscopic
  probabilities'', Physical Review A 67~(2) (2003) 026101.

\bibitem{balian1987equiprobability}
R.~Balian, N.~Balazs, Equiprobability, inference, and entropy in quantum
  theory, Annals of Physics 179~(1) (1987) 97--144.

\bibitem{balian2006microphysics}
R.~Balian, From Microphysics to Macrophysics: methods and applications of
  statistical physics, Vol. 1 and 2, Springer, Berlin, 2006.

\bibitem{vonmises1957probability}
R.~Von~Mises, Probability, statistics, and truth (Dover Publications, New York, 1957).

\bibitem{birnbaum1940properties}
Z.~Birnbaum, H.~S. Zuckerman, On the properties of a collective, American
  Journal of Mathematics 62~(1) (1940) 787--791.

\bibitem{daneri1962quantum}
A.~Daneri, A.~Loinger, G.~M. Prosperi, Quantum theory of measurement and
  ergodicity conditions, Nuclear physics 33 (1962) 297--319.

\bibitem{jaynes1957information1}
E.~T. Jaynes, Information theory and statistical mechanics, Physical Review
  106~(4) (1957) 620.

\bibitem{jaynes1957information2}
E.~T. Jaynes, Information theory and statistical mechanics. ii, Physical Review
  108~(2) (1957) 171.

\bibitem{bell1975wavepacket}
J.~S. Bell, On wave packet reduction in the Coleman-Hepp model, Helvetica
  Physica Acta 48 (1975) 93--98.

\bibitem{vanHove1962}
L.~Van~Hove, Master equation and approach to equilibrium for quantum systems,
  in: E.~D.~G. Cohen (Ed.), Fundamental problems in statistical mechanics,
  North-Holland, Amsterdam, 1962, pp. 157--172.

\bibitem{vanKampen1962}
N.~G. Van~Kampen, Fundamental problems in statistical mechanics of irreversible
  processes, in: E.~D.~G. Cohen (Ed.), Fundamental problems in statistical
  mechanics, North-Holland, Amsterdam, 1962, pp. 173--202.

\bibitem{landau1980statistical}
L.~D. Landau, E.~Lifshitz, Statistical physics, part I, Pergamon, Oxford, 1980,
  Chapter 1.

\bibitem{bander1996approach}
M.~Bander, Approach to microcanonical equilibrium for nonisolated systems,
  arXiv preprint cond-mat/9609218.

\bibitem{auffeves2014contexts}
A.~Auff{\`e}ves, P.~Grangier, Contexts, systems and modalities: a new ontology
  for quantum mechanics, Foundations of Physics 46 (2016) 121--137.

\bibitem{auffeves2015simple}
A.~Auff{\`e}ves, P.~Grangier, A simple derivation of Born's rule with and
  without Gleason's theorem, arXiv preprint arXiv:1505.01369.

\bibitem{gleason1957measures}
A.~M. Gleason, Measures on the closed subspaces of a Hilbert space, Journal of
  mathematics and mechanics 6~(6) (1957) 885--893.

\bibitem{busch2003quantum}
P.~Busch, Quantum states and generalized observables: a simple proof of
  Gleason's theorem, Physical Review Letters 91~(12) (2003) 120403.

\bibitem{caves2004gleason}
C.~M. Caves, C.~A. Fuchs, K.~K. Manne, J.~M. Renes, Gleason-type derivations of
  the quantum probability rule for generalized measurements, Foundations of
  Physics 34~(2) (2004) 193--209.

\bibitem{hess2005bell}
K.~Hess, W.~Philipp, Bell's theorem: Critique of proofs with and without
  inequalities, in: AIP Conference Proceedings, Vol. 750, 2005, p. 150.

\bibitem{khrennikov2007bell}
A.~Khrennikov, Bell's inequality: Nonlocalty, ``death of reality'', or
  incompatibility of random variables?, in: Quantum theory: Reconsideration of
  Foundations 4, Vol. 962, AIP Publishing, 2007, pp. 121--131.

\bibitem{hess2009possible}
K.~Hess, K.~Michielsen, H.~D. Raedt, Possible experience: From Boole to Bell,
Europhysics Letters 87~(6) (2009) 60007.

\bibitem{nieuwenhuizen2011contextuality}
T.~M. Nieuwenhuizen, Is the contextuality loophole fatal for the derivation of
  Bell inequalities?  Foundations of Physics 41~(3) (2011) 580--591.

\bibitem{kupczynski2015bell}
M.~Kupczynski, Bell inequalities, experimental protocols and contextuality,
  Foundations of Physics 45~(7) (2015) 735--753.

\bibitem{greenberger1989going}
D.~M. Greenberger, M.~A. Horne, A.~Zeilinger, Going beyond Bell's theorem, in:
  M.~Kafatos (Ed.), Bell's theorem, quantum theory and conceptions of the
  universe, Springer, 1989, pp. 69--72.


\end{thebibliography}

\end{document}